\documentclass{article}
\usepackage[utf8]{inputenc}
\usepackage{authblk}

\usepackage[english]{babel}
\usepackage{latexsym}
\usepackage{indentfirst,cite}
\usepackage{amsmath}
\usepackage{amssymb}
  \usepackage{pdfpages}
 \usepackage{bm}
 \usepackage{comment}
\usepackage{graphicx} 
\usepackage{amsfonts,soul,mathrsfs}
\usepackage{hyperref}

\usepackage[normalem]{ulem}

\newcommand{\be}{\begin{equation}}
 \newcommand{\ee}{\end{equation}}
\newcommand{\bear}{\be\begin{array}}
\newcommand{\bea}{\begin{eqnarray}}
\newcommand{\eea}{\end{eqnarray}}

\newcommand{\bp}{{\bf p}}
\newcommand{\br}{{\bf r}}

\newcommand{\bk}{{\bf k}}

\newcommand{\la}{\langle}
\newcommand{\ra}{\rangle}

\usepackage{xcolor}

\newcommand{\dst}{\displaystyle}
\newcommand{\fr}[2]{\frac{{\dst #1}}{{\dst #2}}}

\begin{document}

\title{Attosecond physics hidden in Cherenkov radiation}
\author{Dmitry Karlovets\thanks{dmitry.karlovets@metalab.ifmo.ru}}
\author{Alisa Chaikovskaia}
\author{Dmitriy Grosman}
\author{Daria Kargina}
\author{Aleksandr Shchepkin}
\author{Georgii Sizykh}
%\author[$1$]{
\affil[]{{\small School of Physics and Engineering, ITMO University, 197101 St.\,Petersburg, Russia}}

\date{\today}

\maketitle

\begin{abstract}
Cherenkov radiation of charged particles moving with superluminal velocities in transparent media is a well-studied phenomenon with a plethora of applications. Its microscopic origins can be traced to the polarization of atomic shells, characterized by time scales in the subfemtosecond range — dynamics that eludes conventional macroscopic treatment. Here we present a theoretical framework for probing the intrinsic dynamics of Cherenkov radiation, unveiling quantum features absent in classical realm and even in a fully quantum theory in momentum space. These features include a finite formation length and spreading time of the photon, the latter becoming negative nearby the Cherenkov angle, a finite flash duration tied to the size of the electron packet, along with a shift in the photon arrival time that can be either positive or negative and necessitates going beyond the far-field approximation. The calculated time scales lie in the attosecond range for the relevant parameters, thus linking this macroscopic phenomenon back to its atomic origins. Finally, we propose that by measuring the duration of the Cherenkov flash one can in principle retrieve the length of the emitting packet, deepening our understanding of quantum coherence effects in photon emission.
\end{abstract}

\section*{Introduction}

Cherenkov radiation (ChR) by charged particles in media \cite{Cherenkov, Vavilov, Frank1937, Ginzburg1940, Sokolov1940, Tamm_Nobel, Frank_Nobel, Frank, NIM, Bolotovsky} is the simplest example of a wide range of phenomena embracing transition radiation, diffraction, Smith-Purcell radiation, and other mechanisms of photon emission \cite{Ginzburg, Mono_DR, Smith, Curcio}. Their common microscopic origin is atomic bremsstrahlung \cite{Korol} due to dynamic polarization of atomic shells by the field of the charge, and the characteristic time scales are femto- and attoseconds, typical for the AC Stark effect \cite{Delone}. Along with numerous applications of classical ChR for particle detection { in acceleration experiments and cosmic rays studies} \cite{Forty_Book}, neutrino telescopes \cite{Koshiba_Nobel, Baikal}, gamma-ray astronomy \cite{Aliu, Depaoli}, and other fields \cite{Bolotovsky}, it has also recently attracted  attention of the biomedical community as a new tool for molecular imaging and therapy in cancer treatment~\cite{Color_imaging, phototherapy, Glaser_2015}. 

{A quantum theory of ChR in momentum space was presented in 1940 by Ginzburg \cite{Ginzburg1940} and Sokolov \cite{Sokolov1940} who predicted a correction to the classical Tamm-Frank result due to quantum recoil, which is negligibly small for the overwhelming majority of applications of ChR. Interest in inherently quantum features of ChR was revived in 2016 \cite{IdoKaminerOAM, Ivanov2016, IdoKaminerNonpertubative} after vortex electrons with quantized orbital angular momentum projection were generated at electron microscopes \cite{Bliokh, IvanovRev}. Some flaws in the initial analysis of Ref.\cite{IdoKaminerOAM} were corrected in Ref.\cite{Ivanov2016}. More recently, it has been argued that the spatiotemporal features of ChR can be connected to the spatial coherence of the emitting charged particle \cite{Karnieli}, whereas generalized measurements of either the final electron or the photon can lead to the generation of a wave packet of the other particle with the needed properties \cite{EPJC}. 

The spatiotemporal characteristics of radiation in the pre-wave zone are neccesary for applications of such close relatives of ChR as transition, diffraction, and Smith-Purcell radiation, for instance, in bunch-length measurements at accelerators \cite{Verzilov, Mono_DR}. The classical theory of the Cherenkov wake fields in the near-field zone has also been developed for beam position monitors and radiation sources at accelerators and free-electron lasers \cite{Cohen, Awake, Floettmann_ChR} and even alternative concepts of the Cherenkov wakefield acceleration have been proposed \cite{Voin}. However, no quantum theory of ChR and its generalizations exists so far beyond the far-field approximation. %\sout{most likely because this approximation is} 
The latter is tightly linked to the conventional momentum-space approach, whereas a spatiotemporal analysis of the photon field at a finite distance from the emitting particle necessitates working in real space and time, and that is always tricky in relativistic quantum theory. Although the flash duration of ChR was first estimated by Frank as early as in 1956 from classical considerations \cite{Frank,NIM}, the microscopic atomic dynamics in the formation of the ChR field remains hidden even in the fully quantum treatment of ChR in momentum space.}

Here we point out that one can access atomic time scales in ChR and in its generalizations by using a quantum theory in phase space where -- similar to quantum optics \cite{Schleich} -- we employ a Wigner function to characterize the emitted photon field. We demonstrate how to probe the field in the pre-wave -- or formation -- zone \cite{Verzilov, Mono_DR, Galletti} in which the partial waves interfere, the Cherenkov cone is not formed yet, and the emitted energy propagates in real space and time as a spreading wave packet. {We find that} our theory establishes a link between the evolving coherence length of the electron packet, the Cherenkov flash duration, and a quantum temporal delay that the photon experiences in medium  and that can only be quantitatively studied beyond the far-field approximation. {This delay falls within the attosecond range, a time scale characteristic of atomic excitation processes explored in attosecond spectroscopy and metrology \cite{Krausz}, and it can be either positive or negative, that is, the photon wave packet can reach the detector on average later or sooner than one can expect from the classical far-field considerations. This temporal detuning} also shares some similarities with tunneling-related time effects, such as those observed in electron ionization~\cite{Dienstbier} and emission~\cite{attoclock}. The quantum shift in the photon arrival time, coupled with the finite flash duration and the spreading dynamics, unveils intricate coherence properties that enable refined temporal control in quantum emission processes, enriching the landscape of quantum optics, ultrafast physics, and of applications of ChR and related phenomena.

\begin{figure}\center
\includegraphics[scale=1.00] {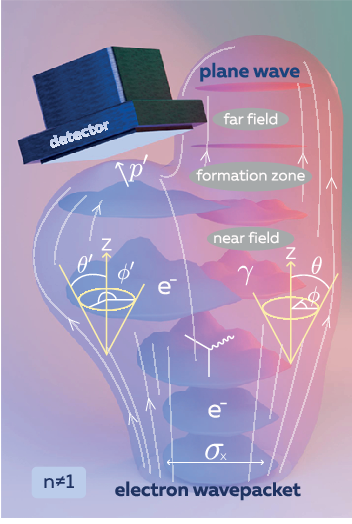}
\caption{\textbf{Spatiotemporal dynamics of the Cherenkov radiation generated by a spreading electron wavepacket.} $\sigma_x$ stands for the initial size of an electron packet (blue), $e$, and $n$ is the refractive index of the medium. {The radiation process begins with the fundamental quantum electrodynamics interaction $e\rightarrow e'+\gamma$, during which} the photon field (pink), $\gamma$, gradually separates from the electron packet within the formation zone and spreads. The electron is detected with a momentum $\bp'=|\bp'|\{\sin\theta'\cos\phi',\sin\theta'\sin\phi',\cos\theta'\}$, and the Cherenkov cone is formed in the far-field only if the electron remains undetected. {In phase-space picture, the photon with the momentum $\bk = n\omega\{\sin\theta\cos\phi,\sin\theta\sin\phi,\cos\theta\}$ can be detected at a finite distance, meaning it does not necessarily reach the far field.
%but is instead located in a region of space centered at the point $\br$ at the moment of time $t$.
}}
    % \caption{Generation of Cherenkov radiation by a  spreading electron packet with an  initial rms size $\sigma_x$ in a dielectric medium with a refractive index $n$. The photon field gradually becomes separated from the electron in the formation zone and spreads. The Cherenkov cone is formed in the far-field only if the electron is not detected. The phase space picture implies that the photon with a momentum $\bk$ is detected  at a finite distance -- that is, not necessarily in the far field -- in a region of space centered at the point $\br$ at the moment of time $t$.}
    \label{Figgen}
\end{figure}

\section*{Results}
\subsection*{Photon emission in phase space}

The system of units $\hbar = c = 1$ is used, the electron velocity is $\beta = u_p/c \equiv u_p$, $m = 0.511$ MeV is the electron mass, $1/m \equiv \hbar/mc = 3.86\times 10^{-11}$ cm {is the reduced Compton wavelength, and the corresponding timescale is} $t_\text{C} = 1/m \equiv \hbar/mc^2 \approx 1.3\times 10^{-21}\,\text{s}$. When an electron emits a photon, the two-particle state within the first order of the perturbation theory  in quantum electrodynamics is $|e',\gamma\ra = (\hat{1} + \hat{S}^{(1)})|\text{in}\ra$, where $|\text{in}\ra = |e_{\text{in}}\ra\otimes|0_{\gamma}\ra$ and $\hat{S}^{(1)} = -ie\int d^4x\, \hat{j}^{\mu}(x)\hat{A}_{\mu}(x)$ \cite{BLP}. Fig.\ref{Figgen} illustrates how the photon  field gradually becomes spatially separated from the electron  packet, spreads, and eventually turns into a plane wave propagating at the angles $\theta,\phi$. If we detect the electron in a plane-wave state $\la \bp',\lambda'|$, $\lambda' = \pm 1/2$, the state of the photon becomes
\bea
& \displaystyle |\gamma\ra = \la \bp',\lambda'|e_{\text{in}}\ra|0_{\gamma}\ra + \sum\limits_{\lambda_\gamma}\int\frac{d^3k}{(2\pi)^3}\, |\bk,\lambda_\gamma\ra\, S_{\text{fi}}.
\label{gammaGM}
\eea
The second term here is a coherent superposition of plane waves with the momenta $\bk$ and the helicity $\lambda_{\gamma} = \pm 1$ and a matrix element is $S_{\text{fi}} = \la \bk,\lambda_\gamma;\bp',\lambda'|\hat{S}^{(1)}|\text{in}\ra$. We treat the incoming electron as a  Gaussian packet with a wave function $f_\text{e}^{(\text{in})}(\bp,\lambda) = \la \bp,\lambda|e_{\text{in}}\ra$, a mean momentum $\la\bp\ra$, an uncertainty $\sigma \ll m$, which means that the rms size -- also called coherence length -- of the packet is much larger than the Compton wavelength, $\sigma_x = 1/\sigma \gg 1/m$. For the moment, we take a simplified model with the packet, spherically symmetric in the laboratory frame, $\sigma_{\perp} = \sigma_{||} \equiv \sigma_x = 1/\sigma$.

Let us define the energy density of the photon field in real space and time, an electric part of which is (Supplementary Notes 1 and 2)
\bea
& \displaystyle \fr{1}{8\pi}\la\gamma|\hat{\bm E}^2({\bf r}, t)|\gamma\ra \to \fr{1}{4\pi}\left|\la 0|\hat{\bm E}({\bf r}, t)|\gamma\ra\right|^2 = \int\frac{d^3k}{(2\pi)^3}\, \mathcal W(\br, \bk, t),
\label{EdensityMod}
\eea
where we have subtracted a contribution of the vacuum energy and $\hat{\bm E}({\bf r}, t)$ is a secondary-quantized electric field operator \cite{Scully}. A contribution of the magnetic field can be written  in a similar way. Here
\bea
& \displaystyle \mathcal W(\br, \bk, t) = \frac{1}{4\pi} \sum\limits_{\lambda_{\gamma},\tilde{\lambda}_{\gamma}}\int\frac{d^3{\tilde k}}{(2\pi)^3}\,\bm E^*_{\tilde{\lambda}_{\gamma}}(\bk - {\tilde \bk}/2)\cdot\bm E_{\lambda_{\gamma}}(\bk+{\tilde \bk}/2)\, e^{-it \left(\omega(\bk+{\tilde \bk}/2)-\omega(\bk-{\tilde \bk}/2)\right) + i\br\cdot{\tilde \bk}}
\label{EW}
\eea
is a Wigner function of the photon,  and $\bm E_{\lambda_{\gamma}}(\bk) = \frac{i\omega \sqrt{4\pi}}{\sqrt{2\omega n^2}}\,{\bm e}_{\bk\lambda_{\gamma}}\sum\limits_{\lambda}\int\frac{d^3p}{(2\pi)^3}\,f_\text{e}^{(\text{in})}(\bp,\lambda) S_{\text{fi}}^{(\text{pw})},\, {\bm e}_{\bk\lambda_{\gamma}}\cdot\bk=0,$ is a positive-frequency component of its electric field. 

According to the standard interpretation \cite{Scully}, the energy density Eq.~(\ref{EdensityMod}) defines probability of detecting a photon in a region of space centered at the point $\br$ at the moment of time $t$. Clearly, the second marginal distribution $\int d^3x\,\mathcal W(\br, \bk, t) \propto |S_{\text{fi}}^{(\text{pw})}|^2$ yields probability to detect a photon with the frequency $\omega$ and a wave vector $\bk,\, |\bk| = n(\omega)\,\omega$, the standard result of the quantum theory in momentum space (see \cite{Ivanov2016}). Therefore, it is this Wigner function Eq.~(\ref{EW}) in phase space that contains all the information on spatial distribution of the photon  energy density at a given distance $\br$ -- also in the near-field zone -- and on its dynamics. Consequently, emission takes place in the pre-wave zone even if the condition of ChR is {\it not} met, but the waves do not constructively interfere to form a cone in the far field. 

We calculate the Wigner function in the paraxial approximation, $\sigma \ll m$, in a medium with {\it weak dispersion}, $\frac{\omega}{n(\omega)}\frac{dn(\omega)}{d\omega} \ll 1$, and represent the tree-level amplitude \cite{BLP, Ivanov2016} as follows:
\bea
S_{\text{fi}}^{(\text{pw})} = |S_{\text{fi}}^{(\text{pw})}|\,e^{i\zeta_{\text{fi}}}.
\eea
Here $|S_{\text{fi}}^{(\text{pw})}|^2$ defines the emission rate in momentum space where the phase \cite{Bliokh, TOTEM, IvanovRev} $\zeta_{\text{fi}}$ does {\it not} contribute to the probability, although it is non-vanishing even in the lowest order of the perturbation theory (see Supplementary Note 5). The result of the calculations is (see Supplementary Note 3)
\bea
\label{Wigner}
& \displaystyle  \mathcal W(\br, \bk, t) \propto \int\limits_0^{\infty}dt'\,\fr{e^{-R^2/R_{\text{eff}}^2(t')}}{G(t')}\,\cos\left(F(t')\right),
\label{WEEpar}
\eea
where $G(t')>0$ is {the function particular form of which is of no concern for the following discussion}, the momentum conservation $\bp = \bp' + \bk$ is implied, and
\bea
& \displaystyle {\bf R} = \br - {\bm u}_p t + (\partial_{\bp} + \partial_{\bk})\zeta_{\text{fi}}(\bp, \lambda_e, \bk, \lambda_{\gamma}). 
\label{Rr}
\eea
Here ${\bm u}_p = \bp/\varepsilon(\bp),\, \varepsilon(\bp) = \sqrt{\bp^2 + m^2},\, {\bm u}_k = \bk/(n |\bk|), |{\bm u}_k| = 1/n$,  $\partial_{\bp} = \partial/\partial \bp$, $F(t') \propto \arctan t'/t_\text{d}$ contains a Gouy phase of the photon connected to its spreading with time $t'$, and $t_\text{d}$ is a diffraction time (see below). 

The  spatio-temporal dependence of the Wigner function Eq.~(\ref{WEEpar}), which is {\it not} everywhere positive  even in the paraxial approximation, is governed by the ratio
\bea
& \displaystyle \fr{R^2}{R_{\text{eff}}^2(t')} =\fr{1}{\sigma_x^2(t')}\Bigg(\underbrace{\fr{[{\bf R}\times ({\bm u}_p - {\bm u}_k)]^2}{({\bm u}_p - {\bm u}_k)^2}}_{\text{finite at $t'=0$}} + \text{terms vanishing at}\, t'= 0\Bigg),
\label{Rexp}
\eea
where $\sigma_x^2(t') = \sigma^{-2}\left(1+(t'/t_\text{d})^2\right)$ is an rms size of the electron packet. When the condition of ChR is met, $u_p > u_k$, the vector ${\bm u}_k - {\bm u}_p$ is directed backwards with respect to the electron velocity ${\bm u}_p$, and dependence of the Wigner function on ${\bf R}$ at small $t'$ vanishes along ${\bm u}_k - {\bm u}_p$, defining {\it the Mach cone} with an angle
\bea
\theta_{\text{Mach}} = \pi - \arcsin\left(\fr{\sin\theta}{n|{\bm u}_k - {\bm u}_p|}\right).
\label{Machth}
\eea
If the electron is detected in a plane-wave state, scattered at the angles $\theta',\phi'$, the radius $R_{\text{eff}}(t')$ depends on the difference $\phi_R - \phi$ between the azimuthal angle of ${\bf R}$ and that of $\bk$, so it is {\it anisotropic}. The azimuthal symmetry of the Mach cone is restored when the electron is {\it not} detected and we integrate Eq.(\ref{WEEpar}) over $\bp'$.

\begin{figure}\center
\includegraphics[scale=0.65] {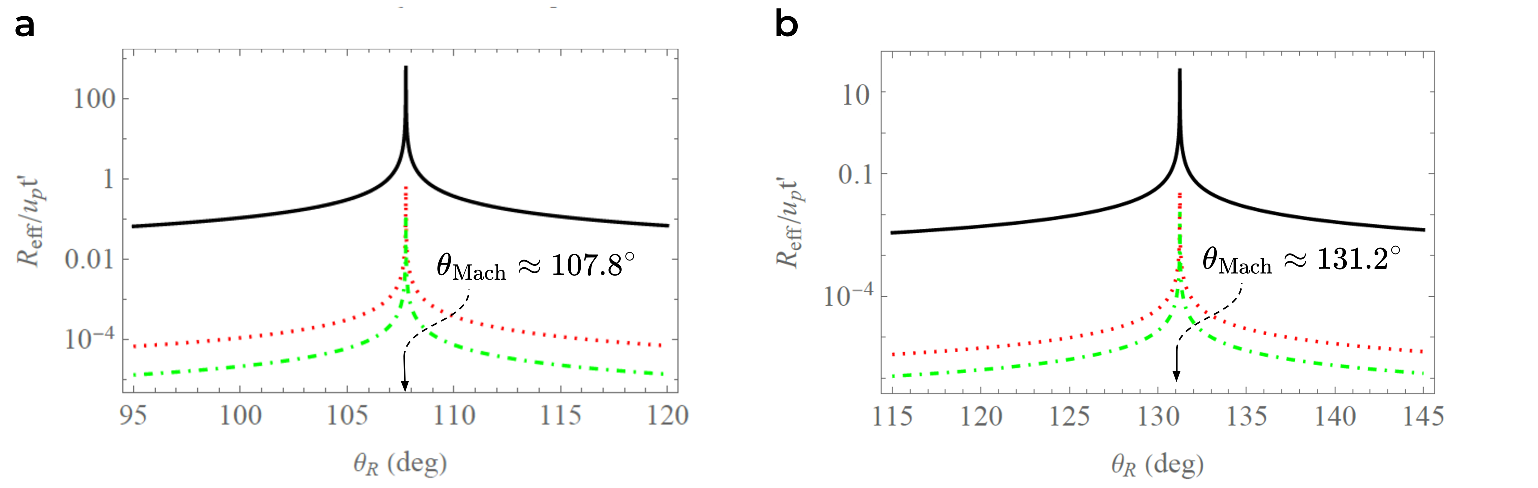}
    \caption{\textbf{The effective correlation radius of the Cherenkov radiation.} The  ratio $R_{\rm{eff}}$ to the distance $u_p t'$ is given for the electron traveling times $t' = 10^7\,t_\text{C}$ (solid black line), $t' = 10^{10}\,t_\text{C}$ (dotted red line) and $t' = 10^{14}\,t_\text{C}$ (dot-dashed green line). At panel \textbf{(a):} $\beta = 0.7\,(\gamma = 1.4), n=1.5,\,\theta = \theta_{\text{Ch.cl.}} = \arccos (1/u_p n) \approx 17.8^\circ, \, \omega = 10^{-6}m, \,\sigma = 10^{-5}m,\,\phi_R-\phi = 0$ deg, $\theta_{\text{Mach}} \approx 107.8^\circ$. At panel \textbf{(b):} $\beta = 0.9999\,(\gamma = 70.7), n=1.33,\, \theta =  \theta_{\text{Ch.cl.}} \approx 41.2^\circ, \omega = 10^{-5}m, \, \sigma = 10^{-4}m,\,\phi_R-\phi = 0$ deg, $\theta_{\text{Mach}} \approx 131.2^\circ$. Nearby the Mach angle $\theta_{\text{Mach}}$, space-time dependence of the Wigner function quickly vanishes within the correlation radius $R < R_{\text{eff}}(t')$, which is a hallmark of the wave zone.}
    \label{FigReffLog}
\end{figure}

\begin{figure}\center
\includegraphics[scale=0.65]{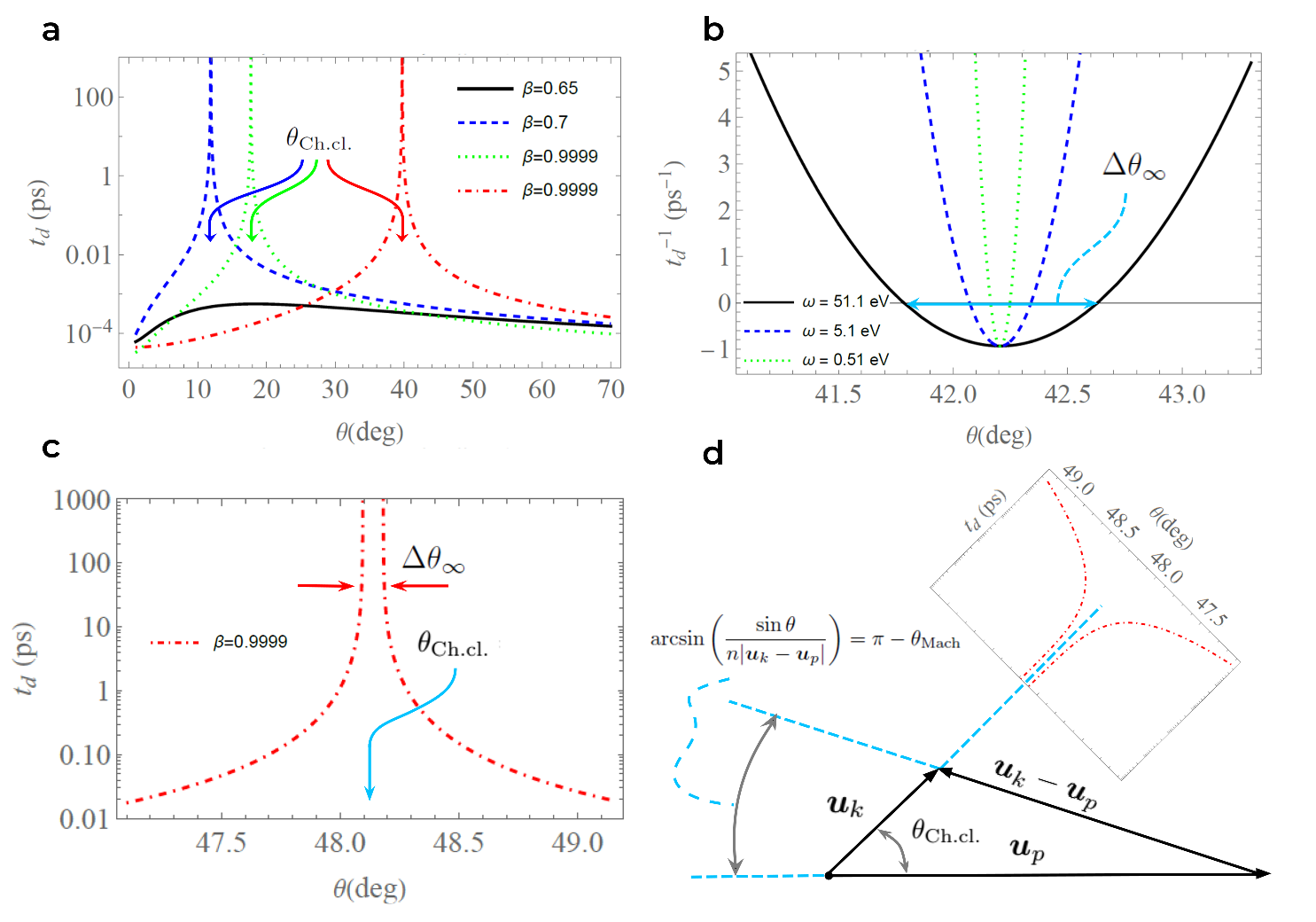}
    \caption{\textbf{The spreading time $t_\text{d}$ of the photon field, angular spans $\Delta \theta_{\infty}$ defining negative areas for $t_\text{d}$ and the Mach cone of the emission.}  Panel \textbf{(a)}: We take $\omega = 0.51$ eV, $\sigma = 10^{-5}$ and the packet width $\sigma_x = 1/\sigma \approx 38$ nm. The spreading time $t_\text{d}$ (in picoseconds) from Eq.~(\ref{td}) is displayed for the electron velocity $\beta = 0.65$ and refractive index $n = 1.46$ (solid black), $\beta = 0.7$ and $n = 1.46$ (dashed blue), $\beta = 0.9999$ and $n = 1.05$ (dotted green) and $\beta = 0.9999$ and $n = 1.3$ (dot-dashed red line).  The sharp maxima are nearby the angle $\theta_{\text{Ch.cl.}} = \arccos 1/u_p n$ as illustrated by the arrows. The Cherenkov condition is {\it not} met for the black line, which is why the photon field quickly spreads during hundreds of attoseconds. Panel \textbf{(b)}: The inverse spreading time is plotted for parameters $n =1.5, \, \beta = 0.9, \, \sigma = 10^{-4}$m and $\omega = 51.1$~eV (solid black), $\omega = 5.1$~eV (dashed blue) and $\omega = 0.51$~eV (dotted green line). In the angular span of a size $\Delta \theta_{\infty}$ between two points given by  Eq.~(\ref{thetainf})  the spreading time turns negative due to the quantum recoil ($\omega/\varepsilon \ne 0$, see Eq.~(\ref{Dth})). Panel \textbf{(c)}: A close-up picture of the spreading time $t_\text{d}$ for parameters $n=1.5, \, \beta = 0.999,\, \omega = 5.1$ eV, $\sigma = 10^{-4}$m, $\theta_{\text{Ch.cl}} \approx 48.1$ deg. Panel \textbf{(d)}: Formation of the Mach cone of the electron emission. The electron moves with velocity ${\bm u}_p$ exceeding the photon velocity ${\bm u}_k$. The photon field produces the Mach cone with an angle $\theta_{\text{Mach}}$ according to Eq.(\ref{Machth}).}
    \label{Figtd}
\end{figure}

At large $t'$, we find $R_{\text{eff}}^2(t') \propto \sigma^2_x(0)\, t'^2/t_\text{d}^2$ and when $R_{\text{eff}}(t') \gg R$ dependence of the Wigner function on $\br$ and $t$ {\it vanishes},  similar to the classical wave zone. In the other limiting case, $R \gg R_{\text{eff}}(t')$,  the integrand in Eq.(\ref{WEEpar}) is exponentially suppressed. So, an effective region where space-time correlation exists is when $t' \lesssim t_\text{d}$, and $R \sim R_{\text{eff}}(t')$ is where it is most pronounced, which is why one can call $R_{\text{eff}}(t')$ {\it the correlation radius}. At $t' \gg t_\text{d}$, both the electron packet and the photon field spread and so there is no longer space-time correlation within the region $R < R_{\text{eff}}(t')$, which is a hallmark of the wave zone. We show in Fig.\ref{FigReffLog} that the correlation radius is orders of magnitude {\it smaller} than the distance $u_p t'$ traveled by the electron during the time interval $t'$ for all the angles $\theta_R$ of ${\bf R}$, except for the Mach angle, $\theta_R \approx \theta_{\text{Mach}}$ (shown in Fig.\ref{Figtd} \textbf{d}).

\subsection*{Spreading time and formation length}

Let us discuss now the spreading time, which is found as
\bea
& \displaystyle t_\text{d} = \frac{2}{\sigma^2}\frac{({\bm u}_p - {\bm u}_k)^2}{\left(1/\omega n^2- \varepsilon^{-1}\right) ({\bm u}_p - {\bm u}_k)^2 + \left(\varepsilon^{-1} - \omega^{-1}\right) \left[{\bm u}_p\times {\bm u}_k\right]^2}.
\label{td}
\eea
where $({\bm u}_p - {\bm u}_k)^2 = n^{-2}+u_p^2-2u_p\cos\theta/n,\ \left[{\bm u}_p\times {\bm u}_k\right]^2 = u_p^2\sin^2\theta/n^2$ with
the $z$ axis directed along the electron partial momentum $\bp$. One can also define {\it the formation length} of radiation as $L_\text{f} = u_p |t_\text{d}|$, which turns to infinity at the Cherenkov angle in the classical Tamm problem \cite{Pafomov, Ginzburg, Mono_DR, Konkov}. 

The diffraction time and the formation length have an extremum either at the angle $\cos\theta_{\text{Ch.cl.}} = 1/u_p n < 1$, when the Cherenkov condition $u_p > 1/n$ is met, or at $\cos\theta = u_p n < 1$ otherwise. The Wigner function and the energy density stay finite in the latter case, but the photon field rapidly spreads (see the black line in Fig.\ref{Figtd} \textbf{a}). Along with the extremum, the time $t_\text{d}$ has two points in which its denominator {\it vanishes} (see Fig.\ref{Figtd} \textbf{b}),
\bea
& \displaystyle \cos\theta_{\infty} \approx \fr{1}{u_p n} \left(1 \mp \sqrt{\fr{\omega}{\varepsilon}}\sqrt{(n^2-1)(u_p^2n^2-1)}\right), 
\label{thetainf}
\eea
where we have kept the first correction due to quantum recoil, which is usually very small \cite{Ivanov2016}, $\omega/\varepsilon \ll 1$. Clearly, the diffraction time can only turn to infinity under the condition of ChR, $u_p n >1$. The angular width between the two points is
\bea
\Delta \theta_{\infty} \approx 2\sqrt{\fr{\omega}{\varepsilon}}\sqrt{n^2-1},
\label{Dth}
\eea
and it vanishes for classical emission with no recoil, $\omega/\varepsilon \to 0$. %, in which case $t_\text{d}$ has only one maximum at $\cos\theta = 1/u_p n$.
 For materials like Al, Si, Be, and Ti, Cherenkov radiation can be observed in the soft X-ray range at the frequencies \cite{Uglov-1, Uglov-2, Konkov} $\omega \approx 72.5$, $100$, $110$, $453.8$ eV, respectively, which for $\varepsilon \sim (5-20) m$ yields $\Delta \theta_{\infty} < 1-2\, \text{deg}$ (cf. Fig.\ref{Figtd} \textbf{c}, \textbf{d}).

Between the above points, the time $t_\text{d}$ becomes \textit{negative} -- see Fig.\ref{Figtd} \textbf{b} -- and the Gouy phase $\arctan t'/t_\text{d}$ changes it sign, as if the electron packet {\it shrinks} during the emission. In the classical regime with no recoil, both the points merge and so $t_\text{d}$ and $L_\text{f}$ turn to infinity at the Cherenkov angle. Indeed, in a vicinity of this angle the spreading time with the recoil kept is
\bea
& \displaystyle t_\text{d}\Big|_{\cos\theta = 1/n u_p} = \frac{2\varepsilon}{\sigma^2}\frac{n^2}{1-n^2} < 0, 
\label{tdCh}
\eea
where $\varepsilon=\gamma m$. We compare this with the spreading time $t_\text{d}^{(\text{e, rest})} = m/\sigma^2$ of an electron packet, which is at rest on average in vacuum \cite{NJP}. In the laboratory frame this time is $\gamma$ times larger, which coincides with $|t_\text{d}|$ up to the factor $2n^2/(n^2-1) > 2$. So, spreading of the photon {\it seems to reverse back} in a vicinity of the Cherenkov direction and it is intimately connected with spreading of the electron packet itself.

The coherence lengths of non-relativistic electrons amount to $\sigma_x(0) \sim 1-10\,\text{nm}$ nearby the standard sources like cathodes of the electron guns in accelerators or electron microscopes \cite{Ehberger, Cho, Latychevskaia, Cho_Oshima, NJP}. These estimates can likewise be obtained by using the emission duration of photo-electrons from a tungsten tip \cite{Dienstbier} for which the measured sub-femtosecond duration yields nanometer-sized packets. Therefore for $\gamma \gtrsim 1-2, n \gtrsim 1$ the electron spreading time is $t_\text{d}^{(e)} \gtrsim 10^{-2}-10\, \text{ps}$, and the time $t_\text{d}$ for the photon is of the same order of magnitude nearby the Cherenkov angle, see Fig.\ref{Figtd}. The time of flight of an electron through a target of a few centimeters in length is roughly $0.1\,\text{ns}$, and therefore spreading of the electron {\it can} be relevant even not far from $\theta_{\text{Ch.cl.}}$, especially  for non-relativistic particles and for large Cherenkov generators employed, for instance, in neutrino telescopes.

 One can use the peculiar behavior of the diffraction time nearby the Cherenkov angle as a means for detecting the quantum recoil in ChR in UV or soft X-ray range. For that, one needs to measure the size of the photon wave packet at different distances close to the electron path -- that is, in the pre-wave zone -- and at different polar angles $\theta$ with an angular resolution of at least $0.1$ deg, which is definitely challenging, but not inconceivable.

\subsection*{Shift of the photon arrival time and flash duration}

Dependence of the Wigner function Eq.(\ref{WEEpar}) on the detection time $t$ comes exclusively from the following envelope:
\bea
& \displaystyle \exp\left\{-\fr{R^2}{R_{\text{eff}}^2(t')}\right\} \propto \exp\left\{-\frac{(t - t_0)^2}{2\sigma_t^2(t')}\right\},\cr & \displaystyle \sigma_t^2(t') = \frac{\sigma^2_x(t')}{2}\frac{({\bm u}_p - {\bm u}_k)^2}{\left[{\bm u}_p\times {\bm u}_k\right]^2}.
\label{tdep}
\eea
Here, natural duration of the Cherenkov flash is defined by $\sigma_t(t')$ and $t_0 = {\bm l}_0\cdot\left(\br + (\partial_{\bp} + \partial_{\bk})\zeta_{\text{fi}}\right)$ is a time instant at which the probability to catch the photon around the point $\br$ is maximized, ${\bm l}_0 = \left[({\bm u}_p - {\bm u}_k)\times\left[{\bm u}_k \times {\bm u}_p\right]\right]/\left[{\bm u}_p\times {\bm u}_k\right]^2$.  We will call the time $t_0$ {\it the mean arrival time}.

One can neglect the term with the phase $\zeta_{\text{fi}}$ in ${\bf R}$ Eq.(\ref{Rr}) in the wave zone  where $|\br - {\bm u}_p t| \gg |(\partial_{\bp} + \partial_{\bk})\zeta_{\text{fi}}|$, and then the Wigner function Eq.(\ref{WEEpar}) and the emitted energy seem to be concentrated in a vicinity of the classical electron trajectory \cite{Bagrov}, $\br \sim {\bm u}_pt$. The detector registers a photon in the far field emitted at $t=0, r_0=0$ by a classical point-like electron at the time instant
\bea
t_{\text{cl.}}^{(\text{far-f.})} = r/u_k = r\,n,
\label{tcl}
\eea
which will be called the classical arrival time. Let us compare this prediction with the above $t_0$, derived quantum mechanically. Orienting the $z$ axis along the electron momentum $\bp$, we find $\bk = n \omega {\bm l},\ {\bm l}=\{\sin\theta\cos\phi,\sin\theta \sin\phi,\cos\theta\},\ r \equiv \br\cdot{\bm l},\ t_{\text{cl.}}^{(\text{far-f.})} = \br\cdot{\bm l}\,n,$ and ${\bm l}_0 = \sin^{-2}\theta\, \left(\left(u_p^{-1} - n \cos{\theta} \right) \bp/|\bp| + \left(n - u_p^{-1}\cos\theta\right) {\bm l}\right)$. In a vicinity of the Cherenkov angle, we have ${\bm l}_0 \to {\bm l}\,n$ and so for $\zeta_{\text{fi}} = 0$ we get $t_0 \to t_{\text{cl.}}^{(\text{far-f.})} = \br\cdot{\bm l}\,n$, in accordance with Eq.(\ref{tcl}).

However, the phase $\zeta_{\text{fi}}$ {\it cannot} be ignored in the formation zone and in the near-field (in the latter case, $\br \approx {\bm u}_pt$, so that ${\bf R} \approx (\partial_{\bp} + \partial_{\bk})\zeta_{\text{fi}}$) and this makes the photon arrival time $t_0$ different from the classical value $t_{\text{cl.}}^{(\text{far-f.})}$. In this regime, we call $t_{\text{cl.}}=\br\cdot{\bm l}_0$ for arbitrary emission angles and so the quantum shift is
\bea
\Delta t = t_0 - t_{\text{cl.}} = {\bm l}_0\cdot (\partial_{\bp} + \partial_{\bk})\zeta_{\text{fi}}.
\label{Dtg}
\eea
 This shift is not necessarily positive and the physical origin of this delay or advance is the electric dipole moment density $\propto e(\partial_{\bp} + \partial_{\bk})\zeta_{\text{fi}}$ \textit{induced in medium} by the field of the electron. We deal with an analogue of the AC Stark effect \cite{Delone} with the atoms being off-resonantly polarized by a broadband spectrum $\Delta \omega$ of pseudo-photons. Similarly to the observed time delays -- positive and negative -- when a laser propagates in a medium \cite{Jordan,Sommer,Negtime}, here we encounter delays induced by the virtual photons, reemitted as real ones. Classically, one can look at this as if the photon was emitted not from a point-like electron, but from a point shifted laterally to the distance $\Delta \rho \sim \beta \gamma \lambda/2\pi$ from the electron trajectory closer or further from the detector, which is a mean free path of the virtual photon \cite{Verzilov, Mono_DR}. Numerically $\Delta \rho/c \sim \beta \gamma \lambda/2\pi c = \beta\gamma/\omega \sim 1\,\text{fs}-100\,\text{fs}$ for photons from IR to UV ranges and $\gamma = \varepsilon/m \lesssim 10$.

\begin{figure}\center
\includegraphics[scale=0.45]{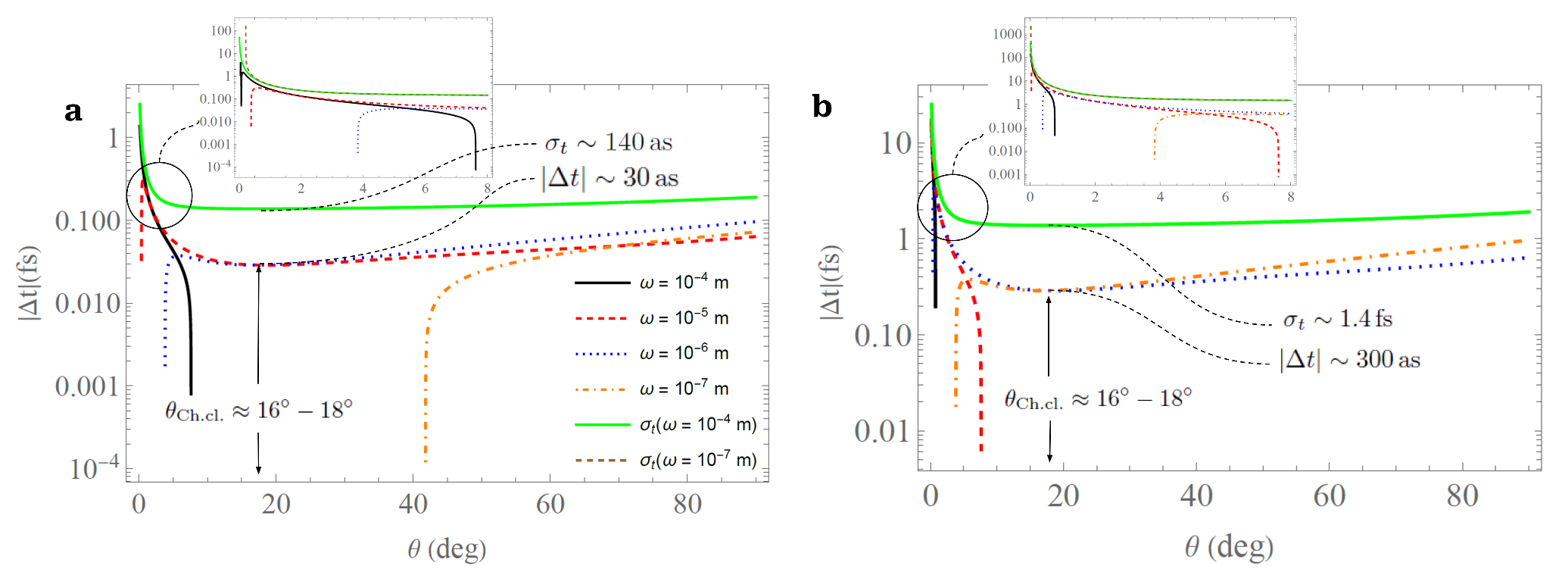}
    \caption{\textbf{The quantum shift of the photon arrival time as a function of the emission angle.} We take the electron energy typical for a transmission electron microscope: $\varepsilon_c \approx 200$ keV, $\beta \approx 0.7$, $p'_{\perp} = 0.99\times p_{\perp},\, p'_z = p_z - k_z \approx 0.98\, m,\, n=1.5$. The electron momentum uncertainty is equal to the transverse momentum: $\sigma=p_{\perp}$. In panel \textbf{(a)}: $p_{\perp} = 10^{-5}\, m,\, 1/p_{\perp} \gtrsim 10\,\text{nm}$; panel \textbf{(b)}: $p_{\perp} = 10^{-6}\, m,\, 1/p_{\perp} \gtrsim 100\,\text{nm}$. The classical Cherenkov flash durations for the emitted photon energies $\omega = 10^{-4}m$ (green solid line) and $10^{-7}m$ (brown dashed line) are compared to the quantum shifts (Eq.~(\ref{Dtg})) for  $\omega = 10^{-4}m$ (black solid), $10^{-5}m$ (red dashed), $10^{-6}m$ (blue dotted) and $10^{-7}m$ (orange dot-dashed line). The behavior at small angles is shown in the inset figures. The quantum shifts are restricted to the regions allowed by the momentum conservation law and vanish outside of them. They stay roughly the same for other values of $p'_{\perp},\, p'_z$ and for ultrarelativistic electrons, $\gamma \gg 1$, though the Cherenkov angle grows.}
    \label{FigDtNonr}
\end{figure}

\begin{figure}\center
\includegraphics[scale=0.47]{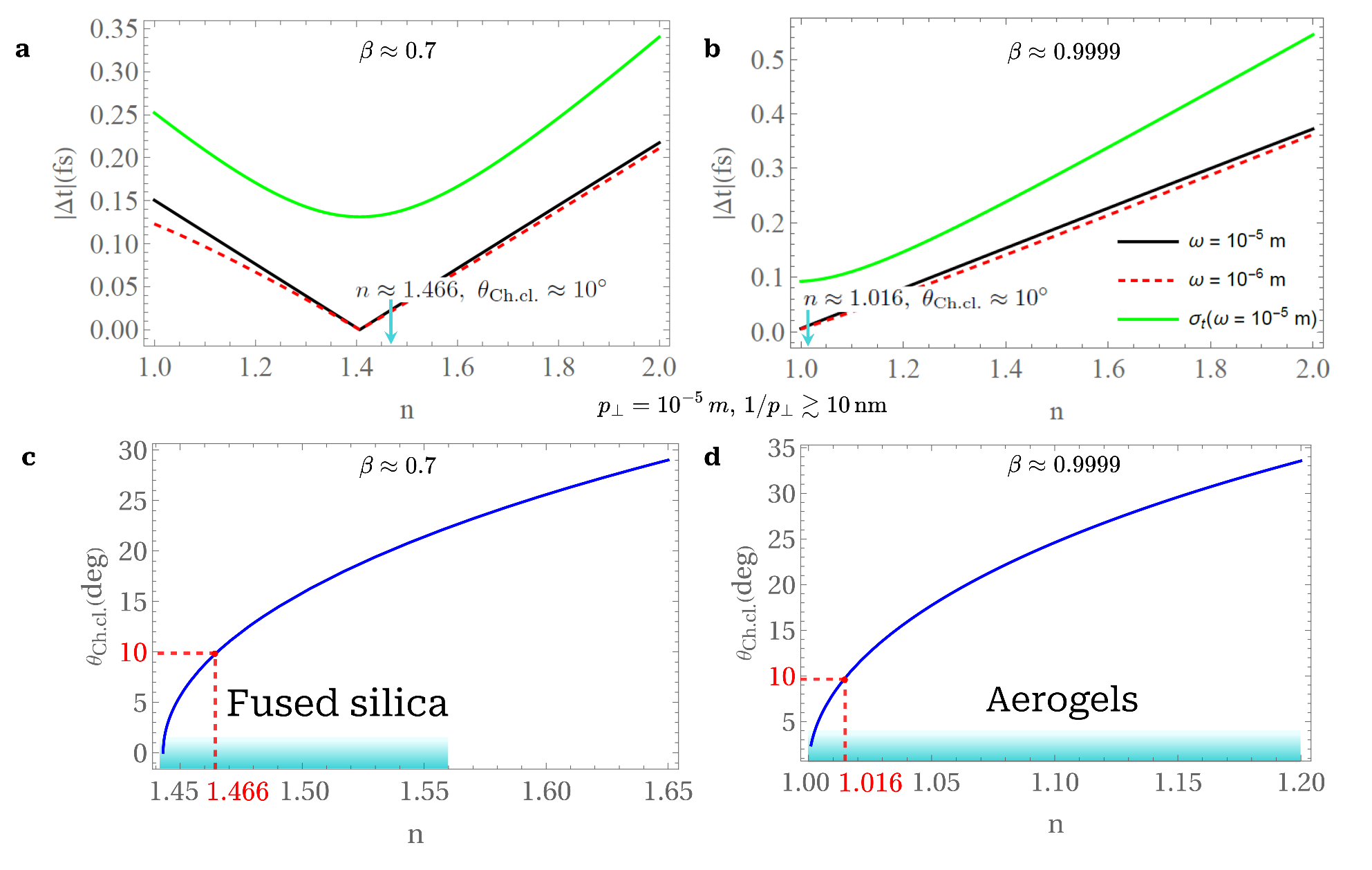}
\caption{\textbf{The quantum shift of the photon arrival time and the classical Cherenkov angle as a function of the refractive index.} Here, the emission angle is chosen to be $\theta = 10$ deg and the electron momentum uncertainty is $\sigma=p_{\perp}$. We distinguish a regime of a transmission electron microscope -- panel \textbf{(a)}: $\beta \approx 0.7\, (\gamma \approx 1)$, and an accelerator regime -- panel \textbf{(b)}: $\beta = 0.9999\, (\gamma \approx 70)$. In both panels \textbf{(a)} and \textbf{(b)} we fix $p'_{\perp} = 0.99\times p_{\perp},\, p'_z = p_z - k_z$, the classical Cherenkov flash duration is shown for $\omega = 10^{-5}m$ (green solid line), whereas the quantum shifts are given for $\omega = 10^{-5}m$ (black solid line) and $10^{-6}m$ (red dashed line). Panels \textbf{(c)} and \textbf{(d)} demonstrate dependence of the classical Cherenkov angle $\theta_{\text{Ch.cl.}} = \arccos(1/\beta n)$ on the refractive index for two types of dielectric media - fused silica and aerogels. The blue shaded areas in \textbf{(c)} and \textbf{(d)} show the boundaries of the refractive indices due to the frequency dispersion of $n(\omega)$ for fused silica or different types of particular chemical solutions for aerogels. The minimal refractive index for which the Cherenkov condition is met in scenario \textbf{(a)} is $n\approx 1.444$, and $\theta_{\text{Ch.cl.}}=10$~deg corresponds to $n\approx 1.466$ within the range of $n$ from panel \textbf{(c)}. Analogously,  in scenario \textbf{(b)}$:\theta_{\text{Ch.cl.}}=10$~deg corresponds to $n\approx 1.016$ within the range of $n$ from panel \textbf{(d)}. }%whereas $n=1.45-1.55$ for fused silica in the given spectral range.}
    \label{FigN2}
\end{figure}

When measuring the quantum shift in the photon arrival time from its classical value, the flash duration $\sigma_t(t')$ is crucial because the deviations can hardly be discerned with $\sigma_t(t') \gg |\Delta t|$ occurring for $t' \gg t_\text{d}$ far from the Cherenkov angle. This duration was estimated classically by Frank \cite{Frank,NIM} to be $\sim 1/\Delta \omega < 1\,\text{ps}$ where $\Delta \omega$ is a frequency interval for which the emission takes place. Quantum estimates from the uncertainty principle {yield roughly the same result both in the macroscopic theory of ChR \cite{Karnieli} and for the microscopic off-resonant AC Stark effect in single atoms} \cite{Delone}, $1/\Delta \omega \sim 0.1-100\,\text{fs}$ for $\Delta \omega \sim 10^{-2}-10$ eV. Our quantum model predicts the following flash duration nearby the Cherenkov angle:
\bea
\sigma_t(t') \to n\,\sigma_x(0)/\sqrt{2},
\eea
because in this case $t_\text{d} \to \infty$. Clearly, only the electron packet's length at the target entrance ($t'=0$) contributes to the flash duration nearby $\theta_{\text{Ch.cl.}}$. For realistic electrons with $\sigma_x(0) \sim 1-100\,\text{nm}$, we find
\bea
 & \displaystyle \sigma_t(0) \sim 10\,\text{as} - 1\,\text{fs}.
\label{durt0}
\eea

In Figures \ref{FigDtNonr} and \ref{FigN2} we show that the typical shifts indeed belong to the attosecond range and that the flash duration is  generally larger than the shift. The electron transverse momenta are chosen to be $p_{\perp} = \sigma \sim (10^{-7}-10^{-4})\,m$ because they correspond to the spatial widths $1/p_{\perp} = \sigma_x(0) \gtrsim 1\, \text{nm} - 1\,\mu\text{m}$, respectively, and we also neglect the spreading. Note that within the pre-wave zone the emission does {\it not} take place only at the Cherenkov angle. The sign of the shift {\it swaps} between the two kinematic scenarios (Supplementary Notes 4 and 5), which is why the absolute value $|\Delta t|$ is shown. Fixing the detector at certain angles $\theta, \phi$ and the distance $\bf r$, one would see that the photons equally probably arrive either later (time delay) than $t_{\text{cl.}}$ or sooner than that (an advance or negative delay), if the electron is {\it not} detected. Note that integration of the Wigner function over $\bp'$ puts the electron momentum to the definite value $\bp' = \bp - \bk$ with subsequent integration over $\bp$ with the Gaussian distribution. If one wishes to catch only the shifts with {\it one particular sign}, one should detect the photon and the electron in coincidence, which is technically more challenging. 

From Figures \ref{FigDtNonr} and \ref{FigN2} we conclude that the difference between the classical flash duration $\sigma_t$ (green line in both figures) and the temporal shift is minimized for angles $\theta \lesssim 10$~deg and media with small refractive indices $n \sim 1.01 - 1.5$, whereas the ratio $\sigma_t/|\Delta t| \lesssim 2$ for all the angles $\theta < 10$~deg. The argument is correct only when spreading is not taken into account because $\sigma_t(t')$ grows with time $t'$ and the shift does not. {For larger angles, $\theta \gtrsim 10$~deg, or for $n\gtrsim2$ the difference between the flash duration (the green line) and the shift increases}. {When trying to detect the quantum shift for the transmission electron microscope (TEM) energies}, $\beta \sim 0.7$, a target made of fused silica with $n \sim 1.44-1.47$ (see, for example, \cite{Tabata2010, Arosa2020}) % $n\approx 1.458$ 
in the optical range can do the job, whereas for ultrarelativistic electrons, $\gamma \gg 1$, aerogels with $n \sim 1.01-1.30$ can be employed (see, for example, \cite{Malitson1965, Bellunato2008}), which are already used as Cherenkov generators \cite{Adachi,sol-gel,SKIF}. The use of Brewster-Cherenkov detectors \cite{Lin} can also come in handy here. {Note, however, that for the TEMs energies there also be transition radiation generated at a boundary of a real target, which propagates at the angles of $\theta \sim \gamma^{-1}$ and can, therefore, interfere with the Cherenkov signal. Although microscopically this type of radiation has a similar origin as ChR and, therefore, similar temporal shifts are likely to take place, one can start with studying pure Cherenkov emission. One of the alternatives would be to employ the so-called diffraction Cherenkov radiation \cite{Konkov, Mono_DR} where an electron moves close to a dielectric target but does not intersect it. In that case, diffraction radiation is generated at the angles $\theta \ll \gamma^{-1}$ and Cherenkov emission can still be well distinguished at $\theta \lesssim 10$ deg.}

If we now go beyond the {simplified} model of the electron packet with $\sigma_{\perp} = \sigma_{||}$ and recall that there is Lorentz contraction, $\sigma_{||} = \gamma^{-1}\sigma_{\perp}$, the flash durations can become $\gamma$ times {\it shorter} than Eq.(\ref{durt0}). Although photon spreading can be safely neglected nearby the Cherenkov angle, spreading of the electron {\it before} entering the target can significantly increase these numbers. According to the quantum dynamics of the electron coherence length $\sigma_x(t')$ -- also called the generalized van Cittert-Zernike theorem \cite{NJP} -- when an electron is released from a photo-gun or a field emitter with rms sizes $\sigma_{\perp} \sim \sigma_{||}$ \cite{Ehberger} of a few nanometers and an energy up to a few tens of eV, it spreads to some tens of nanometers at the distance of 1 $\mu\text{m}$ and it reaches micrometer sizes already at 1 mm from the source. Clearly, electron acceleration -- say, in an RF cavity -- mitigates the spreading rate, but a realistic estimate of the packet length $\sigma_{||}$ at the target boundary remains {\it unknown}  and it is not usually controlled. {Therefore, measuring the flash duration in a vicinity of the ChR angle, one can retrieve the length of the electron packet at the target entrance, a complementary approach to interferometry that is reminiscent of bunch length measurements via classical coherence \cite{Curcio}.}

In practice, achieving attosecond flash durations requires {nanometer-sized electron packets} generating radiation with no spreading before the target, just after the emission from a cathode. A vacuum gap of even a few millimeters after the source {and before the target} would yield micrometer-sized electron packets entering the target, which results in picosecond flash durations {or in subpicosecond ones with acceleration to a few MeV in the gap due to the above factor $\gamma^{-1}$}, measurable by streak cameras. State-of-the-art Cherenkov counters at accelerators have picosecond time resolution \cite{SKIF, Shaikh}, the subfemtosecond resolution can be achieved at X-ray free-electron lasers \cite{Guo, Duris} or with the frequency-resolved optical gating \cite{Trebino}, whereas it is nanoseconds for Cherenkov telescopes in gamma-ray astronomy \cite{Depaoli} where spreading effects are notable. Attosecond photon pulses -- including twisted photons with orbital angular momentum -- are usually obtained through high-harmonic generation in the extreme-ultraviolet and even soft X-ray ranges \cite{Heras, Geneaux}, enabling vortex electrons generation via photoionization \cite{Geneaux}. The ChR can be a source of twisted photons \cite{Ivanov2016, EPJC}, also in the soft X-ray range, and shaping the spatial and temporal profiles of the electron wave function offers refined tuning of phase-space profile of the photons. {In particular, it seems feasible to generate photon wave packets with the given spatial profile, vorticity, and the energies up to the soft X-ray range with the attosecond and even sub-attosecond durations.}

\section*{Conclusions}

The developed quantum framework provides visualization of the emitted photon energy in phase space and, in particular, in real space and time, including the near-field zone. This can be desirable for estimating the Cherenkov wakefields in accelerator chambers as well as for biomedical applications of ChR in radiotherapy \cite{phototherapy, Glaser_2015} where only the classical Tamm-Frank theory in the far field has been used so far, apparently overestimating the radiation intensity in the formation zone. {We have predicted several spatiotemporal features of the quantum ChR, absent within the conventional momentum-space analysis limited to the far-field approximation. In particular, the Cherenkov flash duration close to the Cherenkov angle turns out to be determined by the electron coherence length upon entering the medium, opening avenues for controlling this duration by selecting packets of the desired length and accelerating them in the low-current regime with no space-charge effects, typical for TEMs. By using the charged particles with the Lorentz factors $\gamma=\varepsilon/m$ from $\gamma\sim 10$ to $\gamma\sim 10^3-10^4$, one can generate attosecond photon flashes and even the sub-attosecond ones if the target is installed close to the particle source. Such short photon pulses can come in handy for various studies in atomic physics, including those of the temporal effects in electron excitation and photoionization. 

Alternatively, measuring photon pulse durations with subpicosecond resolution can provide a technique for determining the lengths of the emitting charged-particle packets. This can be used to control the lengths of the wave packets in electron microscopes and accelerator facilities (linacs or storage rings with electrons, protons, or ions),  as well as of ultrarelativistic charged particles coming from the cosmos. Naturally, ChR serves as an exemplary case within a broader class of media-induced emission phenomena where attosecond time scales are accessible via phase-space analysis.}

\

{\bf Acknowledgment}. We are grateful to D.\,Glazov, I.\,Ivanov, V.\,Serbo, I.\,Pavlov, and especially to A.\,Tishchenko for useful discussions and suggestions, and also to Maria Zhuravleva for the help with the 3D picture. The studies of general temporal features of Cherenkov radiation are supported by the Ministry of Science and Higher Education of the Russian Federation (Project FSER-2025-0012). Those of the photon field at a finite distance are supported by the Russian Science Foundation (Project No. 23-62-10026 https://rscf.ru/en/project/23-62-10026/). The analysis of the quantum shift is supported by the Government of the Russian Federation through the ITMO Fellowship and Professorship Program. The studies on emission of an electron packet in phase space (by D. Karlovets, D. Grosman, A. Shchepkin, and G. Sizykh) are supported by the Foundation for the Advancement of Theoretical Physics and Mathematics “BASIS”. 

\

{\bf Author contributions}. All authors contributed extensively to this work. Dm.K. conceived the idea, developed the theoretical framework, wrote the manuscript with input from all the authors and acquired the funding. D.G., G.S., A.C., A.Sh. and D.K. checked the calculations and interpreted the results in detail. D.G., A.C., and G.S. assisted with the figures, while A.C., Dm.K., and G.S. refined the presentation of results and theoretical details.

\

{\bf Data Availability}. Data sharing not applicable to this article as no datasets were generated or analysed during the current study.

\

{\bf Competing interests}. The authors declare no competing interests.

\bibliographystyle{h-physrev}
\bibliography{refs}

\end{document}

\title{Supplementary Information \\ to Attosecond physics hidden in Cherenkov radiation}
\author{Dmitry~Karlovets}
\author{Alisa~Chaikovskaia}
\author{Dmitriy~Grosman}
\author{Daria~Kargina}
\author{Aleksandr~Shchepkin}
\author{Georgii~Sizykh}
%\author[$1$]{
\affil[]{{\small School of Physics and Engineering, ITMO University, 197101 St.\,Petersburg, Russia}}

%\date{\today}

\maketitle

\section*{Supplementary Note 1: The photon evolved state}

Let us describe photon emission in QED with a final state consisting of an electron and a photon with Cherenkov radiation (ChR) in a transparent medium being specific example. A bipartite evolved state is obtained by acting on the initial state by an evolution operator within the first order of the perturbation theory \cite{BLP, PS}, 
\bea
\hat{S} \approx  \hat{1} + \hat{S}^{(1)} = \hat{1} -ie\int d^4x\, \hat{j}^{\mu}(x)\hat{A}_{\mu}(x),
\eea
where the integration over time spans from $t_i=-\infty$ to $t_f=+\infty$. So, the evolved state is 
\bea
|e',\gamma\ra = \left(\hat{1} + \hat{S}^{(1)}\right)|\text{in}\ra, 
\eea
where $|\text{in}\ra = |e_{\text{in}}\ra\otimes|0_{\gamma}\ra$. One can insert a unity operator $\hat{1}_{e\gamma}$ on the two-particle space with the complete set being the plane-wave states with momenta $\bp', \bk$ and helicities $\lambda'=\pm 1/2,\lambda_\gamma = \pm 1$. So that 
\bea
\displaystyle |e',\gamma\ra = |\text{in}\ra + \sum\limits_{\lambda'=\pm 1/2,\lambda_\gamma=\pm 1} \int\frac{d^3k}{(2\pi)^3}\frac{d^3p'}{(2\pi)^3}\, |\bp',\lambda'\ra\otimes|\bk,\lambda_\gamma\ra\, S_\text{fi}^{(1)}.
\label{2partWFPW}
\eea
If we now project the final electron state to a bra $\la f_\text{e}^{(\text{det})}| = \la\bp',\lambda'|$, the evolved state of the final photon alone becomes
\bea
& \displaystyle |\gamma\ra = \la\bp',\lambda'|e_{\text{in}}\ra|0_{\gamma}\ra + \sum\limits_{\lambda_\gamma}\int\frac{d^3k}{(2\pi)^3}\, |\bk,\lambda_\gamma\ra\, S_\text{fi}^{(1)},\cr
& \displaystyle S_\text{fi}^{(1)} \equiv  S_\text{fi}^{(1)}(\bp',\lambda',\bk,\lambda_{\gamma}) = \la \bk,\lambda_\gamma;\bp',\lambda'|\hat{S}^{(1)}|\text{in}\ra.
\label{1partPhGenPW}
\eea
The electron momentum states are on-shell with the energy $\varepsilon'_e = \sqrt{m^2 + (\bp')^2}$. The incoming electron is described as a Gaussian packet with the following wave function:
\bea
& \displaystyle f_\text{e}^{(\text{in})} (\bp,\lambda) = \la \bp,\lambda|e_{\text{in}}\ra = \delta_{\lambda,\lambda_e} \left(\frac{2\sqrt{\pi}}{\sigma}\right)^{3/2} \exp\left\{-\frac{(\bp - \la\bp\ra)^2}{2\sigma^2}\right\},
\label{WPm}
\eea
with $\sigma \ll m$ being the momentum uncertainty, so that $\sigma_x = 1/\sigma \gg 1/m \equiv \hbar/mc \approx 3.86\times 10^{-11}$ cm is the electron Compton wavelength. For simplicity, we employ first the model with a symmetric packet in the laboratory frame, $\sigma_{\perp} = \sigma_{||} \equiv \sigma_x = 1/\sigma$. In the main manuscript, we discuss possible changes of the predicted effects in a more realistic model with a packet, symmetric in the rest frame with $\la\bp\ra =0$, which experiences Lorentz contraction in the laboratory frame, $\sigma_{||} = \gamma^{-1}\sigma_{\perp}$, where $\gamma = \sqrt{m^2+\la\bp\ra^2}/m$.

\section*{Supplementary Note 2: Spatial energy density and the Wigner function}

Let us define the following Hermitian field operators:
\bea
&  \displaystyle \hat{\bm A}({\bf R}, t) = \sum\limits_{\lambda_{\gamma}=\pm 1} \int\frac{d^3k}{(2\pi)^3} \left({\bm A}_{{\bm k}\lambda_{\gamma}}({\bf R}, t)\, \hat{c}_{{\bm k}\lambda_{\gamma}} + \text{h.c.}\right),\cr
&  \displaystyle \hat{\bm E}({\bf R}, t) = - \frac{\partial \hat{\bm A}({\bf R}, t)}{\partial t}= \sum\limits_{\lambda_{\gamma}=\pm 1} \int\frac{d^3k}{(2\pi)^3}\, i\omega\left({\bm A}_{{\bm k}\lambda_{\gamma}}({\bf R}, t)\, \hat{c}_{{\bm k}\lambda_{\gamma}} - \text{h.c.}\right),\cr
&  \displaystyle \hat{\bm H}({\bf R}, t) = \nabla\times \hat{\bm A}({\bf R}, t) = \sum\limits_{\lambda_{\gamma}=\pm 1} \int\frac{d^3k}{(2\pi)^3}\, i\bk\times\left({\bm A}_{{\bm k}\lambda_{\gamma}}({\bf R}, t)\, \hat{c}_{{\bm k}\lambda_{\gamma}} - \text{h.c.}\right)
\label{FOp}
\eea
where $\hat{c}_{{\bm k}\lambda_{\gamma}}$ are the annihilation operators and
\bea
&  \displaystyle {\bm A}_{{\bm k}\lambda_{\gamma}}({\bf R}, t) = \frac{\sqrt{4\pi}}{\sqrt{2\omega}}\, {\bm e}_{{\bm k}\lambda_{\gamma}}\, e^{-i\omega t + i{\bm k}\cdot{\bf R}},
\eea
where ${\bm e}_{{\bm k}\lambda_{\gamma}}\cdot\bk = 0,\ {\bm e}_{{\bm k}\lambda_{\gamma}}\cdot {\bm e}_{{\bm k}\lambda'_{\gamma}}^* = \delta_{\lambda_{\gamma}\lambda'_{\gamma}}$. Let us also define the following averages (cf. \cite{Scully}):
\bea
&  \displaystyle \la 0|\hat{\bm A}({\bf R}, t)|\gamma\ra = \sum\limits_{\lambda_{\gamma}} \int\frac{d^3k}{(2\pi)^3} {\bm A}_{{\bm k}\lambda_{\gamma}}({\bf R}, t)\,S_\text{fi}^{(1)}(\bk,\lambda_{\gamma}),\cr 
&  \displaystyle \la 0|\hat{\bm E}({\bf R}, t)|\gamma\ra = \sum\limits_{\lambda_{\gamma}} \int\frac{d^3k}{(2\pi)^3} i\omega {\bm A}_{{\bm k}\lambda_{\gamma}}({\bf R}, t)\,S_\text{fi}^{(1)}(\bk,\lambda_{\gamma}),\cr 
&  \displaystyle \la 0|\hat{\bm H}({\bf R}, t)|\gamma\ra = \sum\limits_{\lambda_{\gamma}} \int\frac{d^3k}{(2\pi)^3} i[\bk \times {\bm A}_{{\bm k}\lambda_{\gamma}}({\bf R}, t)]\,S_\text{fi}^{(1)}(\bk,\lambda_{\gamma}).
\label{Fav}
\eea
The diagonal part $\hat{S} - \hat{S}^{(1)} = \hat{1}$ does not contribute to Eq.(\ref{Fav}).

A spatial observable in this problem is \textit{the energy density}, an electric part of which is
\bea
&  \displaystyle \la\gamma|\hat{\bm E}^2({\bf R}, t)|\gamma\ra = \la\gamma|\gamma\ra \sum\limits_{\lambda_{\gamma}}\int\frac{d^3k}{(2\pi)^3}\, \omega^2 |{\bm A}_{{\bm k}\lambda_{\gamma}}({\bf R}, t)|^2  + \cr 
&  \displaystyle + 2 \sum\limits_{\lambda_{\gamma}\lambda'_{\gamma}}\int\frac{d^3k}{(2\pi)^3}\frac{d^3k'}{(2\pi)^3}\,\omega\omega' {\bm A}_{{\bm k}\lambda_{\gamma}}({\bf R}, t)\cdot {\bm A}^*_{{\bm k'}\lambda'_{\gamma}}({\bf R}, t)\, S_\text{fi}^{(1)}(\bk,\lambda_{\gamma})\left(S_\text{fi}^{(1)}(\bk',\lambda'_{\gamma})\right)^*
% \label{FOp}
\eea
where $|{\bm A}_{{\bm k}\lambda_{\gamma}}({\bf R}, t)|^2 = 2\pi/\omega$, so the first term
\bea
\varepsilon_0 = \la\gamma|\gamma\ra \sum\limits_{\lambda_{\gamma}}\int\frac{d^3k}{(2\pi)^3}\, \omega^2 |{\bm A}_{{\bm k}\lambda_{\gamma}}({\bf R}, t)|^2 = \la\gamma|\gamma\ra \sum\limits_{\lambda_{\gamma}}\int\frac{d^3k}{(2\pi)^3}\, 2\pi\omega
\eea
\textit{diverges} and should be associated with the vacuum energy. The diagonal term $\hat{S} - \hat{S}^{(1)} = \hat{1}$ only contributes to the factor $\la\gamma|\gamma\ra$. The finite contribution is 
\bea
&  \displaystyle \la\gamma|\hat{\bm E}^2({\bf R}, t)|\gamma\ra - \varepsilon_0 = 2 \left|\la 0|\hat{\bm E}({\bf R}, t)|\gamma\ra\right|^2.
% \label{FOp}
\eea
The magnetic counterpart looks as follows:
\bea
& \displaystyle \la\gamma|\hat{\bm H}^2({\bf R}, t)|\gamma\ra = \la\gamma|\gamma\ra \sum\limits_{\lambda_{\gamma}}\int\frac{d^3k}{(2\pi)^3}\, \bk^2 |{\bm A}_{{\bm k}\lambda_{\gamma}}({\bf R}, t)|^2 + \cr 
& \displaystyle + 2 \sum\limits_{\lambda_{\gamma}\lambda'_{\gamma}}\int\frac{d^3k}{(2\pi)^3}\frac{d^3k'}{(2\pi)^3}\,[\bk\times {\bm A}_{{\bm k}\lambda_{\gamma}}({\bf R}, t)]\cdot[\bk'\times {\bm A}^*_{{\bm k'}\lambda'_{\gamma}}({\bf R}, t)]\, S_\text{fi}^{(1)}(\bk,\lambda_{\gamma})\left(S_\text{fi}^{(1)}(\bk',\lambda'_{\gamma})\right)^* = \cr
&  \displaystyle = \varepsilon_0 + 2\left|\la 0|\hat{\bm H}({\bf R}, t)|\gamma\ra\right|^2.
% \label{FOp}
\eea

Thus, a finite part of the spatial energy density is obtained as
\bea
\mathcal W(\br, t) = \frac{1}{8\pi}\la\gamma|\hat{\bm E}^2({\bf R}, t)+\hat{\bm H}^2({\bf R}, t)|\gamma\ra -\frac{\varepsilon_0}{4\pi} = \frac{1}{4\pi} \left(\left|\la 0|\hat{\bm E}({\bf R}, t)|\gamma\ra\right|^2+\left|\la 0|\hat{\bm H}({\bf R}, t)|\gamma\ra\right|^2\right),
\label{E2H2}
\eea
and it can be interpreted as a probability to detect the emitted photon in a region of space-time centered at the point $(\br, t)$, whereas the electron is jointly detected as a plane wave with the quantum numbers $\bp',\lambda'$.

Now as a next step we rewrite the electric part as follows:
\bea
& \displaystyle \frac{1}{4\pi} \left|\la 0|\hat{\bm E}({\bf R}, t)|\gamma\ra\right|^2 
=  \frac{1}{4\pi} \left|\sum\limits_{\lambda_{\gamma}}\int\frac{d^3k}{(2\pi)^3}\,\bm E_{\lambda_{\gamma}}(\bk)\, e^{-ikx}\right|^2 = \cr & \displaystyle = \frac{1}{4\pi} \sum\limits_{\lambda_{\gamma},\tilde{\lambda}_{\gamma}}\int\frac{d^3k}{(2\pi)^3}\frac{d^3 {\tilde k}}{(2\pi)^3}\,\bm E^*_{\tilde{\lambda}_{\gamma}}(\bk - {\tilde \bk}/2)\cdot\bm E_{\lambda_{\gamma}}(\bk+{\tilde \bk}/2)\, e^{-it \left(\omega(\bk+{\tilde \bk}/2)-\omega(\bk-{\tilde \bk}/2)\right) + i\br\cdot{\tilde \bk}} \equiv \cr
& \displaystyle \equiv \int\frac{d^3k}{(2\pi)^3}\, \mathcal W(\br, \bk, t),
\label{EdensityMod}
\eea
where
\bea
& \displaystyle \bm E_{\lambda_{\gamma}}(\bk) = \frac{i\omega \sqrt{4\pi}}{\sqrt{2\omega n^2}}\,{\bm e}_{\bk\lambda_{\gamma}}\sum\limits_{\lambda}\int\frac{d^3p}{(2\pi)^3}\,f_\text{e}^{(\text{in})}(\bp,\lambda) S_\text{fi}^{(1)}(\bp, \lambda, \bk,\lambda_{\gamma})
\label{EMod}
\eea
is a positive-frequency component of the electric field of the evolved state from Eq.(\ref{Fav}) and
\bea
& \displaystyle \mathcal W(\br, \bk, t) = \frac{1}{4\pi} \sum\limits_{\lambda_{\gamma},\tilde{\lambda}_{\gamma}}\int\frac{d^3{\tilde k}}{(2\pi)^3}\,\bm E^*_{\tilde{\lambda}_{\gamma}}(\bk - {\tilde \bk}/2)\cdot\bm E_{\lambda_{\gamma}}(\bk+{\tilde \bk}/2)\, e^{-it \left(\omega(\bk+{\tilde \bk}/2)-\omega(\bk-{\tilde \bk}/2)\right) + i\br\cdot{\tilde \bk}},
\label{EW}
\eea
is a Wigner function of the photon evolved state in phase space (see, for instance, \cite{Schleich}). Here $f_\text{e}^{(\text{in})}(\bp,\lambda)$ is the wave function of the incoming electron from Eq.(\ref{WPm}), and we have used the normalization of the photon potential in a transparent medium with a refractive index $n(\omega)>1$, see \cite{Ivanov2016}.

Eq.(\ref{EdensityMod}) shows that one marginal distribution of this Wigner function yields the energy density of the photon field in real space and time. The other marginal distribution (the connection between the matrices $S_\text{fi}, T_\text{fi}$, and $M_\text{fi}$ is given in \cite{BLP, PS})
\bea
& \displaystyle \int d^3x\,\mathcal W(\br, \bk, t) = \frac{\omega}{2n^2} \left|\sum\limits_{\lambda}\int\frac{d^3p}{(2\pi)^3}\,f_\text{e}^{(\text{in})}(\bp,\lambda) S_\text{fi}(\bp, \lambda, \bk, \lambda_{\gamma})\right|^2 = \cr 
& \displaystyle = \frac{\omega}{2n^2} (2\pi)^2 \frac{T}{2\pi} \delta(\varepsilon(\bp)-\varepsilon'(\bp')-\omega(\bk))\, \frac{4\pi}{2\omega(\bk) n^2(\omega(\bk)) 2\varepsilon(\bp) 2\varepsilon'(\bp')}\,\left|\sum\limits_{\lambda}f_\text{e}^{(\text{in})}(\bp,\lambda)\,M_\text{fi}(\bp, \bk,\lambda,\lambda_{\gamma})\right|^2_{\bp=\bp'+\bk}
\label{drW}
\eea
yields probability to detect the photon with the frequency $\omega$ and the wave vector $\bk$ (here $T \to \infty$ is a very long period of time \cite{BLP,PS}), that is, the result of the quantum theory of ChR in momentum space \cite{Ivanov2016}. It does {\it not} depends on a phase of the complex amplitude $M_\text{fi}$, even if the incoming electron is a wave packet. When the latter is a plane wave with the momentum $\la\bp\ra$ and the helicity $\lambda_e$, we have $f_\text{e}^{(\text{in})}(\bp,\lambda) \to \text{const}\,\delta_{\lambda\lambda_e}\,\delta(\bp - \la\bp\ra)$ and Eq.(\ref{drW}) reproduces the standard probability of the plane-wave approximation, 
\bea
\int d^3x\,\mathcal W(\br, \bk, t) \propto \left|S_\text{fi}(\la\bp\ra, \lambda_e, \bk,\lambda_{\gamma})\right|^2.
\eea

To understand what new information -- compared to the momentum space -- the Wigner function in phase space and the marginal distribution (\ref{EdensityMod}) bring about, we write the complex amplitude as
\bea
 & \displaystyle M_\text{fi}({\bm p}, {\bm k},\lambda_e, \lambda_{\gamma}) = \left|M_\text{fi}({\bm p}, {\bm k}, \lambda_e, \lambda_{\gamma})\right|\,\exp\left\{i\zeta_\text{fi}({\bm p}, {\bm k},\lambda_e,\lambda_{\gamma})\right\}
\eea
where $\zeta_\text{fi}(\bp, \bk, \lambda_e,\lambda_{\gamma})$ is a dynamic phase (see, for instance, \cite{JHEP,TOTEM,PRD}). Importantly, this phase \textit{is non-vanishing} even at the tree level -- see Sec.5. The Wigner function of the photon field in phase space is defined by the following \textit{master integral} $\mathcal W_p(\br, \bp, \bk, t)$:
\bea
& \displaystyle \mathcal W(\br, \bk, t) = (2\pi)^5\sum\limits_{\lambda_{\gamma},\tilde{\lambda}_{\gamma}}\sum\limits_{\lambda\tilde{\lambda}} \int\frac{d^3p}{(2\pi)^3}\,\delta(\bp - \bp' - \bk) \int\frac{d^3\tilde{k}}{(2\pi)^3}\, \sqrt{\frac{\omega(\bk+{\tilde \bk}/2)\omega(\bk-{\tilde \bk}/2)}{2n^2\left(\omega(\bk+{\tilde \bk}/2)\right)2n^2\left(\omega(\bk-{\tilde \bk}/2)\right)}}\cr
& \displaystyle \times {\bm e}^*_{\bk-{\tilde \bk}/2,\tilde{\lambda}_{\gamma}}\cdot{\bm e}_{\bk+{\tilde \bk}/2,\lambda_{\gamma}} \left(f_\text{e}^{(\text{in})}(\bp-{\tilde \bk}/2,\tilde{\lambda})\right)^* f_\text{e}^{(\text{in})}(\bp+{\tilde \bk}/2,\lambda)\,\delta\left(\varepsilon(\bp+{\tilde \bk}/2)-\varepsilon' - \omega(\bk + {\tilde \bk}/2)\right)\cr
& \displaystyle \times \delta\left(\varepsilon(\bp-{\tilde \bk}/2)-\varepsilon' - \omega(\bk - {\tilde \bk}/2)\right)  T_\text{fi}\left(\bp+{\tilde \bk}/2,\lambda,\bk+{\tilde \bk}/2,\lambda_{\gamma}\right) T_\text{fi}^*\left(\bp-{\tilde \bk}/2,\tilde{\lambda},\bk-{\tilde \bk}/2,\tilde{\lambda}_{\gamma}\right)\cr
& \displaystyle \times e^{-it \left(\omega(\bk+{\tilde \bk}/2)-\omega(\bk-{\tilde \bk}/2)\right) + i\br\cdot{\tilde \bk}} \equiv \cr
& \displaystyle \equiv \int\frac{d^3p}{(2\pi)^3}\,(2\pi)^3 \delta(\bp - \bp' - \bk)\, \mathcal W_p(\br, \bp, \bk, t) = \mathcal W_p(\br, \bp = \bp' + \bk, \bk, t),
\label{WEE}
\eea
so that $\mathcal W(\br, \bk, t) = \mathcal W_p(\br, \bp = \bp' + \bk, \bk, t)$.
Note that no approximations have been made so far. The product
\bea
& \displaystyle f_\text{e}^{(\text{in})}(\bp + {\tilde \bk}/2,\lambda)\left(f_\text{e}^{(\text{in})}(\bp - {\tilde \bk}/2,\tilde{\lambda})\right)^* = \cr
& \displaystyle = \delta_{\lambda,\lambda_e}\delta_{\tilde{\lambda},\lambda_e} \left(\frac{2\sqrt{\pi}}{\sigma}\right)^{3} \exp\left\{-\frac{(\bp - \la\bp\ra)^2}{\sigma^2} -\left(\frac{{\tilde \bk}}{2\sigma}\right)^2\right\}
\eea
is also Gaussian.

If the final electron is \textit{not measured}, which is often the case for ChR, the spatio-temporal distribution of the emitted energy is obtained by tracing out the electron quantum numbers,
\bea
& \displaystyle \sum\limits_{\lambda'}\int\frac{d^3p'}{(2\pi)^3}\,\frac{1}{4\pi} \left|\la 0|\hat{\bm E}({\bf R}, t)|\gamma\ra\right|^2 
=  \sum\limits_{\lambda'} \int\frac{d^3p'}{(2\pi)^3} \frac{d^3k}{(2\pi)^3}\, \mathcal W(\br, \bk, t) = \cr 
& \displaystyle = \sum\limits_{\lambda'} \int\frac{d^3p}{(2\pi)^3}\frac{d^3k}{(2\pi)^3}\, \mathcal W_p(\br,\bp,\bk,t)\Big|_{\bp' = \bp - \bk}.
\label{EdEInt}
\eea
Thus, by using the phase space formalism, one can trace the dynamics of the photon field formation, spreading, and interference in real space and time, which is unattainable even in the fully quantum theory in momentum space. Moreover, this intrinsic dynamics turns out to be closely tied to the coherence length of the electron packet and to the phase $\zeta_\text{fi}$ of the amplitude. Analogously to quantum optics \cite{Schleich}, we deal with two marginal distributions, complementary to each other, only one of which is commonly used in the photon emission analysis.

\section*{Supplementary Note 3: Paraxial Wigner function}

In calculating the integral over $\tilde{\bk}$ in Eq.(\ref{WEE}), we make the paraxial approximation in which $\sigma \ll m$ and we neglect the terms $\mathcal O\left(\tilde {\bk}\right)$ in the amplitude but keep those of $\mathcal O\left(\tilde {\bk}^2\right)$ in the phase. The resultant integral over $\tilde {\bk}$ is Gaussian. The calculations are very similar to those of the quasi-classical approximation in relativistic quantum mechanics \cite{Bagrov} and there are regions in phase space -- analogous to the well-known turning points -- where this approximation fails to work. In practice, this region lies within very small emission angles, $\theta \sim \sigma/m < 10^{-3}$, which are of no practical interest for studies of ChR.

The linear $\bk$-terms only come to the pre-exponential factor due to
\bea
{\bm e}^*_{\bk-{\tilde \bk}/2,\tilde{\lambda}_{\gamma}}\cdot{\bm e}_{\bk+{\tilde \bk}/2,\lambda_{\gamma}} = \delta_{\tilde{\lambda}_{\gamma}\lambda_{\gamma}} + \frac{{\tilde k}_i}{2}\left({\bm e}^*_{\bk,\tilde{\lambda}_{\gamma}}\cdot\frac{\partial {\bm e}_{\bk,\lambda_{\gamma}}}{\partial k_i} - {\bm e}_{\bk,\lambda_{\gamma}}\cdot\frac{\partial {\bm e}^*_{\bk,\tilde{\lambda}_{\gamma}}}{\partial k_i}\right) + \mathcal O({\tilde k}^2)
\label{eexpan}
\eea
The similar expansion of the amplitudes at $\tilde{\lambda}_{\gamma}=\lambda_{\gamma}, \tilde{\lambda}=\lambda=\lambda_e$ yields
\bea
 & \displaystyle M_\text{fi}(\bp + {\tilde \bk}/2, \bk + {\tilde \bk}/2,\lambda_e,\lambda_{\gamma})  M^*_{fi}(\bp - {\tilde \bk}/2, \bk -{\tilde \bk}/2,\lambda_e,\lambda_{\gamma}) = \cr   & \displaystyle \left(\left|M_\text{fi}(\bp, \bk,\lambda_e,\lambda_{\gamma})\right|^2 + \mathcal O ({\tilde \bk}^2) \right) \exp\left\{i{\tilde \bk}\cdot \left(\partial_{\bp} + \partial_{\bk}\right)\zeta_\text{fi}(\bp, \bk,\lambda_e,\lambda_{\gamma}) + \mathcal O ({\tilde \bk}^3)\right\}.
\eea
Neglecting the linear correction in Eq.(\ref{eexpan}), we need to calculate the following master integral:
\bea
& \displaystyle \mathcal W_p(\br, \bp, \bk, t) = (2\pi)^2\sqrt{4\pi} \sum\limits_{\lambda_{\gamma}}\frac{|M_\text{fi}(\bp,\lambda_e,\bk,\lambda_{\gamma})|^2}{(2n^2(\bk))^2 2\varepsilon' 2\varepsilon(\bp)}\, \int\frac{d^3{\tilde k}}{(2\pi)^3}\frac{dt'}{2\pi}\frac{d\tau}{2\pi}\, \left(f_\text{e}^{(\text{in})}(\bp-{\tilde \bk}/2)\right)^* f_\text{e}^{(\text{in})}(\bp+{\tilde \bk}/2) \cr
& \displaystyle \times \exp\left\{it'(\varepsilon(\bp) - \varepsilon' - \omega(\bk)) + i{\tilde \bk}\cdot(\br - {\bm u}_p t + (\partial_{\bp} + \partial_{\bk})\zeta_\text{fi} + \tau ({\bm u}_p - {\bm u}_k)) + it' \frac{1}{2}\frac{{\tilde k}_i}{2}\frac{{\tilde k}_j}{2} \left(\partial^2_{ij}\varepsilon - \partial^2_{ij}\omega\right)\right\},
\label{MI}
\eea
where we have taken the equality ${\tilde \bk}\cdot {\bm u}_p = {\tilde \bk}\cdot {\bm u}_k$ into account, which holds within the paraxial approximation. Here
\bea
& \displaystyle 
{\bm u}_p = \frac{\partial \varepsilon(\bp)}{\partial\bp} = \fr{\bp}{\varepsilon(\bp)},\, {\varepsilon(\bp)} = \sqrt{m^2+\bp^2},\ {\bm u}_k = \frac{\partial \omega(\bk)}{\partial\bk},\cr
& \displaystyle \partial^2_{ij}\varepsilon(\bp) \equiv \frac{\partial^2\varepsilon(\bp)}{\partial \bp_i\partial\bp_j} = \frac{1}{\varepsilon(\bp)} (\delta_{ij} - ({\bm u}_p)_i({\bm u}_p)_j).
\eea
We take a medium with {\it weak dispersion} from now on, for which 
\bea
\frac{\omega}{n(\omega)}\frac{dn(\omega)}{d\omega} \ll 1,
\eea
and we find (recall that $\bk^2 = n^2(\omega) \omega^2$)
\bea
& \displaystyle {\bm u}_k =\frac{\partial\omega}{\partial\bk} \approx \frac{\bk}{n^2 \omega} = \frac{\bk/|\bk|}{n}.\cr
& \displaystyle \partial^2_{ij}\omega \equiv \frac{\partial^2\omega}{\partial\bk_i\partial\bk_j} \approx \frac{1}{n^2 \omega} \left(\delta_{ij} - \frac{k_ik_j}{\bk^2}\right).
\eea
Clearly, the photon group velocity in the medium, $|{\bm u}_k| = 1/n < 1$ as $n>1$. 

Taking the incoming packet from Eq.(\ref{WPm}), we get
\bea
& \displaystyle \mathcal W_p(\br, \bp, \bk, t) = (2\pi)^2 \sqrt{4\pi}\sum\limits_{\lambda_{\gamma}} \left(\frac{2\sqrt{\pi}}{\sigma}\right)^3 \frac{|M_\text{fi}(\bp,\lambda_e,\bk,\lambda_{\gamma})|^2}{(2n^2(\bk))^2 2\varepsilon' 2\varepsilon(\bp)}\,\exp\left\{- \frac{(\bp-\la\bp\ra)^2}{\sigma^2}\right\} \cr
& \displaystyle\times \int\frac{d^3{\tilde k}}{(2\pi)^3}\frac{dt'}{2\pi}\frac{d\tau}{2\pi}\, \exp\left\{it'(\varepsilon(\bp) - \varepsilon' - \omega(\bk)) -\mathcal {\bm A}\cdot {\tilde \bk} - \frac{1}{2}{\tilde k}_i{\tilde k}_j B_{ij}\right\},
\label{MII}
\eea
where we find 
\bea
& \displaystyle \mathcal {\bm A}(t,\tau) = -i \Big(\br - {\bm u}_p t + (\partial_{\bp} + \partial_{\bk})\zeta_\text{fi} + \tau ({\bm u}_p - {\bm u}_k))\Big),\cr
& \displaystyle B_{ij}(t') = \delta_{ij}\, \frac{1}{2\sigma^2} - \frac{it'}{4}\left(\partial^2_{ij}\varepsilon - \partial^2_{ij}\omega\right) \approx
\delta_{ij} \Big(\frac{1}{2\sigma^2} + \frac{it'}{4}\Big(\frac{1}{\omega n^2} - \frac{1}{\varepsilon}\Big)\Big) + \frac{it'}{4}\Big(\frac{1}{\varepsilon} - \frac{1}{\omega}\Big)({\bm u}_p)_i({\bm u}_p)_j.
\eea
We evaluate the Gaussian integral as
\bea
& \displaystyle \int\frac{d^3{\tilde k}}{(2\pi)^3}\, \exp\left\{-\mathcal {\bm A}\cdot {\tilde \bk} - \frac{1}{2}{\tilde k}_i{\tilde k}_jB_{ij}\right\} = (2\pi)^{-3/2}\frac{1}{\sqrt{\det B}}\,\exp\left\{\frac{1}{2} B_{ij}^{-1}\mathcal A_i \mathcal A_j\right\},
\label{Gauss}
\eea
where
\bea
& \displaystyle \det B = \eta^2 (\eta + \chi\, {\bm u}_p^2),\ B_{ij}^{-1} = \eta^{-1}\delta_{ij} - \frac{\chi}{\eta(\eta + \chi\, {\bm u}_p^2)}({\bm u}_p)_i({\bm u}_p)_j,\cr
& \displaystyle \eta(t') = \frac{1}{2\sigma^2} + \frac{it'}{4}\Big(\frac{1}{\omega n^2} - \frac{1}{\varepsilon}\Big),\ \chi(t') = \frac{it'}{4}\Big(\frac{1}{\varepsilon} - \frac{1}{\omega}\Big).
\eea
The integral over $\tau$ is also Gaussian and evaluated as follows:
\bea
& \displaystyle \int\limits_{-\infty}^{+\infty}\frac{d\tau}{2\pi}\,\exp\left\{\frac{1}{2}B^{-1}_{ij}\mathcal A_i \mathcal A_j\right\} = \frac{1}{\sqrt{2\pi}}\, \sqrt{\frac{\eta + \chi {\bm u}_p^2}{({\bm u}_p - {\bm u}_k)^2 + \frac{\chi}{\eta} [{\bm u}_p\times {\bm u}_k]^2}}\cr
& \displaystyle \times \exp\left\{-\frac{1}{2\eta} \frac{[{\bf R}\times({\bm u}_p - {\bm u}_k)]^2 + \frac{\chi}{\eta} ({\bm u}_p\cdot \left[{\bf R} \times {\bm u}_k\right])^2}{({\bm u}_p - {\bm u}_k)^2 + \frac{\chi}{\eta} \left[{\bm u}_p\times {\bm u}_k\right]^2}\right\},
\eea
where
\bea
& \displaystyle {\bf R} = \br - {\bm u}_p t + (\partial_{\bp} + \partial_{\bk})\zeta_\text{fi}(\bp, \lambda_e, \bk, \lambda_{\gamma}) \equiv \cr
& \displaystyle \equiv \{X,Y,Z\} = R\{\sin\theta_R\cos\phi_R, \sin\theta_R\sin\phi_R, \cos\theta_R\}.
\label{Rr}
\eea

We can rewrite the pre-exponential factor as follows: 
\bea
& \displaystyle  \sqrt{\frac{\eta + \chi\, {\bm u}_p^2}{\det B \left(({\bm u}_p - {\bm u}_k)^2 + \frac{\chi}{\eta} [{\bm u}_p\times {\bm u}_k]^2\right)}} = \sqrt{\frac{1}{\eta \left(\eta({\bm u}_p - {\bm u}_k)^2 + \chi [{\bm u}_p\times {\bm u}_k]^2\right)}} \equiv \frac{1}{G(t')}\, \exp\left\{-\frac{i}{2}\, g(t')\right\},\cr & \displaystyle G(t') = \Bigg[\left(\frac{1}{(2\sigma^2)^2} + \left(\frac{t'}{4}\right)^2 \left(\frac{1}{\omega n^2} - \frac{1}{\varepsilon}\right)^2\right)\cr
& \displaystyle \times
\left(\frac{\left({\bm u}_p - {\bm u}_k\right)^4}{(2\sigma^2)^2}  + \left(\frac{t'}{4}\right)^2 \left(\left(\frac{1}{\omega n^2} - \frac{1}{\varepsilon}\right)\left({\bm u}_p - {\bm u}_k\right)^2 + \left(\frac{1}{\varepsilon} - \frac{1}{\omega}\right)\left[{\bm u}_p\times {\bm u}_k\right]^2\right)^2\right)\Bigg]^{1/4},\cr
& \displaystyle g(t') = \arctan \frac{t'}{8\sigma^2}\frac{2\left(\frac{1}{\omega n^2} - \frac{1}{\varepsilon}\right)\left({\bm u}_p - {\bm u}_k\right)^2 + \left(\frac{1}{\varepsilon} - \frac{1}{\omega}\right)\left[{\bm u}_p\times {\bm u}_k\right]^2}{\frac{\left({\bm u}_p - {\bm u}_k\right)^2}{(2\sigma^2)^2}  - \left(\frac{t'}{4}\right)^2\left(\frac{1}{\omega n^2} - \frac{1}{\varepsilon}\right) \left(\left(\frac{1}{\omega n^2} - \frac{1}{\varepsilon}\right)\left({\bm u}_p - {\bm u}_k\right)^2 + \left(\frac{1}{\varepsilon} - \frac{1}{\omega}\right)\left[{\bm u}_p\times {\bm u}_k\right]^2\right)}. % \equiv \cr
%& \displaystyle \equiv \arctan \frac{\mp t'}{t_\text{d}},
\label{G1}
\eea
One can alternatively represent the l.h.s of Eq.(\ref{G1}) in terms of two Gouy phases as follows:
\bea
& \displaystyle  g(t') = g_1(t') + g_2(t')
\eea
where
\bea
& \displaystyle g_1(t') = \arctan\fr{t'}{t_\text{d}},\ t_\text{d} = \frac{2}{\sigma^2}\frac{({\bm u}_p - {\bm u}_k)^2}{\left(\frac{1}{\omega n^2}- \frac{1}{\varepsilon}\right) ({\bm u}_p - {\bm u}_k)^2 + \left(\frac{1}{\varepsilon}- \frac{1}{\omega}\right) \left[{\bm u}_p\times {\bm u}_k\right]^2},
\cr
& \displaystyle
g_2(t') = \arctan\fr{t'}{\tilde{t}_d},\ \tilde{t}_d = \fr{2}{\sigma^2}\fr{1}{\fr{1}{\omega n^2}-\fr{1}{\varepsilon}} = \fr{2}{\sigma^2}\fr{\omega n^2}{1-n^2\omega/\varepsilon}.
\label{gtp}
\eea
Note that $\tilde{t}_d = t_\text{d}(\theta=0)$ where the photon emission angle $\theta$ is counted from the electron partial momentum $\bp$. Therefore
\bea
& \displaystyle G(t') = \fr{\left|{\bm u}_p - {\bm u}_k\right|}{2\sigma^2}\left[\left(1+(t'/t_\text{d})^2\right)\left(1+(t'/{\tilde t}_d)^2\right)\right]^{1/4},\cr
& \displaystyle \sqrt{\frac{\eta + \chi\, {\bm u}_p^2}{\det B \left(({\bm u}_p - {\bm u}_k)^2 + \frac{\chi}{\eta} [{\bm u}_p\times {\bm u}_k]^2\right)}} = \fr{2\sigma^2}{\left|{\bm u}_p - {\bm u}_k\right| \left[\left(1+(t'/t_\text{d})^2\right)\left(1+(t'/{\tilde t}_d)^2\right)\right]^{1/4}}\cr
& \displaystyle \times \exp\left\{-\fr{i}{2} \left(\arctan\fr{t'}{t_\text{d}} + \arctan\fr{t'}{{\tilde t}_d}\right)\right\}.
\label{Gouy2}
\eea

So the master integral within the paraxial approximation becomes
\bea
\label{Wigner}
& \displaystyle  \mathcal W_p(\br, \bp, \bk, t) = \left(\frac{2\sqrt{\pi}}{\sigma}\right)^3 \,\frac{\sqrt{4\pi}}{(2n^2(\bk))^2 2\varepsilon' 2\varepsilon(\bp)}\,\sum\limits_{\lambda_{\gamma}} |M_\text{fi}(\bp,\lambda_e,\bk,\lambda_{\gamma})|^2\,\exp\left\{- \frac{(\bp-\la\bp\ra)^2}{\sigma^2}\right\} \cr
& \displaystyle \times \int\limits_{-\infty}^{+\infty}\frac{dt'}{2\pi}\,\frac{1}{G(t')}\exp\Bigg\{it'(\varepsilon(\bp) - \varepsilon' - \omega(\bk)) -\frac{i}{2}\, g(t') -\cr
& \qquad \qquad \qquad \qquad\qquad \qquad\qquad \displaystyle - \frac{1}{2\eta(t')} \frac{\eta(t')[{\bf R}\times({\bm u}_p - {\bm u}_k)]^2 + \chi(t') ({\bf R}\cdot \left[{\bm u}_p \times {\bm u}_k\right])^2}{\eta(t')({\bm u}_p - {\bm u}_k)^2 + \chi(t') \left[{\bm u}_p\times {\bm u}_k\right]^2}\Bigg\}.
\label{WEEpar}
\eea
This expression is not applicable at the very small angles $\theta \lesssim \sigma/m \ll 1$ where $\sigma/m$ is usually smaller than $10^{-3}$ meaning that the spatial coherence length of the electron packet $\sigma_x = 1/\sigma$ is larger than a few angstroms.

The exponent in Eq.(\ref{WEEpar}) can be presented as follows:
\bea
& \displaystyle - \frac{1}{2\eta(t')} \frac{\eta(t')[{\bf R}\times({\bm u}_p - {\bm u}_k)]^2 + \chi(t') ({\bf R}\cdot \left[{\bm u}_p \times {\bm u}_k\right])^2}{\eta(t')({\bm u}_p - {\bm u}_k)^2 + \chi(t') \left[{\bm u}_p\times {\bm u}_k\right]^2} = \cr
& \displaystyle = -\sigma^2\fr{1-it'/\tilde{t}_d}{1+(t'/\tilde{t}_d)^2}\fr{1-it'/t_\text{d}}{1+(t'/t_\text{d})^2} \left(\fr{[{\bf R}\times ({\bm u}_p - {\bm u}_k)]^2}{({\bm u}_p - {\bm u}_k)^2} \left(1 + \fr{it'}{\tilde{t}_d}\right) + it' \fr{\sigma^2}{2}\Big(\frac{1}{\varepsilon} - \frac{1}{\omega}\Big)\fr{({\bf R}\cdot \left[{\bm u}_p \times {\bm u}_k\right])^2}{({\bm u}_p - {\bm u}_k)^2}\right),%\cr
%& \displaystyle t_\text{d}({\bf R}) = \fr{2}{\sigma^2}\fr{({\bm u}_p - {\bm u}_k)^2}{\Big(\frac{1}{\omega n^2} - \frac{1}{\varepsilon}\Big)[{\bf R}\times ({\bm u}_p - {\bm u}_k)]^2 + \Big(\frac{1}{\varepsilon} - \frac{1}{\omega}\Big)({\bf R}\cdot \left[{\bm u}_p \times {\bm u}_k\right])^2}.
\label{explhs}
\eea
The real part of this is
\bea
& \displaystyle \text{Re}(\ref{explhs}) = -\fr{1}{\sigma_x^2(t')}\left(\underbrace{\fr{[{\bf R}\times ({\bm u}_p - {\bm u}_k)]^2}{({\bm u}_p - {\bm u}_k)^2}}_{\text{finite at $t'=0$}} + \underbrace{\fr{(t')^2}{\tau_d^2 (1 + (t'/\tilde{t}_d)^2)}\fr{({\bf R}\cdot \left[{\bm u}_p \times {\bm u}_k\right])^2}{({\bm u}_p - {\bm u}_k)^2}}_{\text{due to spreading at $t'\ne 0$}}\right) \equiv - \fr{R^2}{R_{\text{eff}}^2(t')},\cr 
& \displaystyle \sigma_x^2(t') = \sigma^{-2}\left(1+(t'/t_\text{d})^2\right),\ \tau_d^2 = \fr{2}{\sigma^2}\fr{t_\text{d}\tilde{t}_d}{\left(\fr{1}{\varepsilon}-\fr{1}{\omega}\right)(t_\text{d} + \tilde{t}_d)},
\label{Rexp}
\eea
The imaginary part
\begin{equation}
    \text{Im}(\ref{explhs}) = \frac{t'}{\sigma^2(t')}\left(\frac{1}{t_\text{d}}\fr{[{\bf R}\times ({\bm u}_p - {\bm u}_k)]^2}{({\bm u}_p - {\bm u}_k)^2} - \frac{\sigma^2}{2}\left(\fr{1}{\varepsilon}-\fr{1}{\omega}\right)\frac{1-t'^2/(t_\text{d}\tilde{t}_d)}{1+t'^2/\tilde{t}_d^2}\fr{({\bf R}\cdot \left[{\bm u}_p \times {\bm u}_k\right])^2}{({\bm u}_p - {\bm u}_k)^2}\right )
\end{equation}
vanishes at $t'=0$. And so
\bea
\label{Wigner}
& \displaystyle  \int\limits_{-\infty}^{+\infty}\frac{dt'}{2\pi}\,\frac{1}{G(t')}\exp\Bigg\{it'(\varepsilon(\bp) - \varepsilon' - \omega(\bk)) -\frac{i}{2}\, g(t') - \frac{1}{2\eta(t')} \frac{\eta(t')[{\bf R}\times({\bm u}_p - {\bm u}_k)]^2 + \chi(t') ({\bf R}\cdot \left[{\bm u}_p \times {\bm u}_k\right])^2}{\eta(t')({\bm u}_p - {\bm u}_k)^2 + \chi(t') \left[{\bm u}_p\times {\bm u}_k\right]^2}\Bigg\} = \cr
& \displaystyle = 2\int\limits_0^{\infty}dt'\,\fr{e^{-R^2/R_{\text{eff}}^2(t')}}{G(t')}\,\cos\left(t'(\varepsilon(\bp) - \varepsilon' - \omega(\bk)) -\frac{1}{2}\, g(t') +  \text{Im}(\ref{explhs})\right),
\label{tpr}
\eea
which is why the master integral and the Wigner function are real but \textit{not everywhere positive}. We employ this expression in the main part of the manuscript.

The paraxial Wigner function implies the momentum conservation law for every partial wave, $\bp = \bp' + \bk$ where $|\bk| = n\, \omega(\bk)$, but in the phase space there is {\it no} corresponding energy conservation for the partial waves, $\varepsilon(\bp) - \varepsilon' - \omega(\bk) \ne 0$, due to spreading and dependence of the integrand in Eq.(\ref{tpr}) on $t'$. As a result, the well-known Cherenkov condition of the momentum space (see, e.g., \cite{Ivanov2016}) does \textit{not} hold within the formation or pre-wave zone,
\bea
\cos\theta \ne \cos\theta_{\text{Ch}} = \frac{1}{\beta n} + \frac{\omega}{2\varepsilon}\,\frac{n^2-1}{\beta n},
\label{Changle}
\eea
which is why the photon field is {\it not} vanishing at the angles different from $\theta_{\text{Ch}}$, a hallmark of the pre-wave zone \cite{Verzilov, Mono_DR}. Likewise, there is no sharp spectral cutoff (see the debates in \cite{IdoKaminerOAM, Ivanov2016})
\bea
\omega \nless \omega_{\text{cut-off}} = 2\varepsilon\,\fr{\beta n - 1}{n^2 - 1},
\eea
even neglecting the dispersion of $n(\omega)$. One can see from Eq.(\ref{WEEpar}) that the common features of the far-field Cherenkov radiation are regained when we neglect the spreading, that is, the dependence on time $t'$ of the terms under the integral in Eq.(\ref{tpr}). This can also be done when the incoming electron is a delocalized plane wave with $\sigma \to 0$ because the Gouy phase vanishes when either $t' \ll t_\text{d}$ or $\sigma \ll m$, and the limit of $G(t'), g(t')$ is the same in both these cases.

\section*{Supplementary Note 4: Transverse momentum conservation}\label{ApTri}

When calculating the above Wigner function and the quantum shift of the photon arrival time, we employ the following representation of the delta function of the transverse momentum conservation in cylindrical coordinates:
\bea
& \displaystyle \delta^{(2)}(\bp_{\perp} - \bp'_{\perp} - \bk_{\perp}) = \cr 
& \displaystyle = \frac{\Theta(p_{\perp}, k_{\perp}, p'_{\perp})}{2\Delta} \Big(\delta\left(\phi' - \phi + \alpha\right)\delta\left(\phi_{\gamma} - \phi - \gamma\right) + \delta\left(\phi' - \phi - \alpha\right)\delta\left(\phi_{\gamma} - \phi + \gamma\right)\Big) = \cr
& \displaystyle = \frac{\Theta(p_{\perp}, k_{\perp}, p'_{\perp})}{2\Delta} \Big(\delta\left(\phi' - \phi_{\gamma} - (\beta-\pi)\right)\delta\left(\phi_{\gamma} - \phi - \gamma\right) + \delta\left(\phi' - \phi_{\gamma} + (\beta-\pi)\right)\delta\left(\phi_{\gamma} - \phi + \gamma\right)\Big),
\label{delta2}
\eea
where $\Delta$ is an area of a triangle with the legs $p_{\perp}, p'_{\perp}, k_{\perp}$ (see \cite{Bliokh}) and the angles $\alpha, \beta, \gamma\, (\alpha + \beta + \gamma = \pi)$,
\bea
& \displaystyle\Delta = \frac{1}{2} p_{\perp}p'_{\perp}\sin\alpha = \frac{1}{2} k_{\perp}p'_{\perp}\sin\beta = \frac{1}{2} p_{\perp}k_{\perp}\sin\gamma,\cr
& \displaystyle 
\alpha = \arccos\left\{\frac{p_{\perp}^2 + (p'_{\perp})^2 - k_{\perp}^2}{2p_{\perp}p'_{\perp}}\right\},\, \beta = \arccos\left\{\frac{(p'_{\perp})^2 + k_{\perp}^2 - p_{\perp}^2}{2k_{\perp}p'_{\perp}}\right\},\, \gamma = \arccos\left\{\frac{p_{\perp}^2 + k_{\perp}^2 - (p'_{\perp})^2}{2k_{\perp}p_{\perp}}\right\},
\eea
and the legs satisfy the triangle rules,
\bea
p_{\perp} \leq k_{\perp} + p'_{\perp},\ p'_{\perp} \leq k_{\perp} + p_{\perp},\ k_{\perp} \leq p_{\perp} + p'_{\perp}.
\label{trirules}
\eea
The function $\Theta(p_{\perp}, k_{\perp}, p'_{\perp})$ in Eq.(\ref{delta2}) equals 1 when these inequalities are simultaneously satisfied and vanishes otherwise. Therefore 
\bea
& \displaystyle\frac{1}{2\Delta} = \frac{1}{k_{\perp}p'_{\perp}\sqrt{1-\cos^2\beta}} = \frac{2}{\sqrt{(k_{\perp} + p'_{\perp} - p_{\perp})(k_{\perp} + p_{\perp} - p'_{\perp})(p_{\perp} + p'_{\perp} - k_{\perp})(k_{\perp} + p'_{\perp} + p_{\perp})}},
\eea
and the singularity at $k_{\perp}\to 0$ is integrable, i.e.
\bea
\int\limits_0^{\infty}\frac{dk_{\perp} k_{\perp}}{2\Delta}\ \text{is finite}.
\eea
The two momentum configurations from Eq.(\ref{delta2}) yield different signs of the shift in the photon arrival time from the main manuscript, keeping the same absolute value -- see Eq.(\ref{zetapm}) below.

\section*{Supplementary Note 5: Helicity amplitudes and the phase}\label{Amp}
The first-order amplitude of emission of a photon by an electron is
\bea
M_\text{fi} = \sqrt{4\pi \alpha_{em}}\, \bar{u}_{\bp'\lambda'}\gamma^{\mu}e^*_{\mu}u_{\bp\lambda} = |M_\text{fi}|\,e^{i\zeta_\text{fi}},
\eea
where $\gamma^{\mu}e^*_{\mu} = - {\bm \gamma}\cdot{\bm e}^*_{\bk\lambda_{\gamma}}$ in the Coulomb gauge, and each bispinor and vector are expanded in the following series:
\bea
& \displaystyle u_{\bp\lambda} = \sum\limits_{\sigma=\pm 1/2} u_{\varepsilon\lambda}^{(\sigma)}\, d^{(1/2)}_{\sigma\lambda}(\theta)\, e^{-i\sigma\phi},\cr
& \displaystyle \bar{u}_{\bp'\lambda'} = \sum\limits_{\sigma'=\pm 1/2} \bar{u}_{\varepsilon'\lambda'}^{(\sigma')}\, d^{(1/2)}_{\sigma'\lambda'}(\theta')\, e^{i\sigma'\phi'},\cr
& \displaystyle {\bm e}_{\bk\lambda_{\gamma}} = \sum\limits_{\sigma_{\gamma}=0,\pm 1} {\bm \chi}^{(\sigma_{\gamma})} \, d^{(1)}_{\sigma_{\gamma}\lambda_{\gamma}}(\theta_{\gamma})\, e^{-i\sigma_{\gamma}\phi_{\gamma}}.
\label{uexp}
\eea
Here $\theta, \phi$ are the angles of the vector $\bp$, whereas $\theta_{\gamma}, \phi_{\gamma}$ are those of $\bk$, and also $\hat{s}_z u_{\varepsilon\lambda}^{(\sigma)} =\sigma u_{\varepsilon\lambda}^{(\sigma)}$, $\left({\bm \chi}^{(\sigma_{\gamma})}\right)^*\cdot {\bm \chi}^{(\sigma'_{\gamma})} = \delta_{\sigma_{\gamma}\sigma'_{\gamma}}$. We also employ the phase convention of Ref.\cite{BLP}, so that $\hat{j}_z u_{\bp\lambda} = 0$ (see details in \cite{EPJC}). The small Wigner functions are
\bea
& \displaystyle d^{(1/2)}_{\sigma\lambda}(\theta) = \delta_{\sigma\lambda} \cos(\theta/2) - 2\sigma\, \delta_{\sigma,-\lambda} \sin(\theta/2),\cr
& \displaystyle  d^{(1)}_{\sigma_{\gamma}\lambda_{\gamma}}(\theta_{\gamma}) = \left\{d^{(1)}_{\lambda_{\gamma}\lambda_{\gamma}} = \cos^2(\theta_{\gamma}/2), d^{(1)}_{-\lambda_{\gamma}\lambda_{\gamma}} = \sin^2(\theta_{\gamma}/2), d^{(1)}_{0\lambda_{\gamma}} = \frac{\lambda_{\gamma}}{\sqrt{2}}\sin(\theta_{\gamma})\right\},\, \lambda_{\gamma} = \pm 1.
\label{uexp1}
\eea
Importantly, the phase $\zeta_\text{fi}$ is only due to the quantized phases of the bispinors and of the photon polarization vector from Eq.(\ref{uexp}), which is why this phase {\it has no classical counterpart}. However, exactly as the phase $e^{i\ell\phi}$ can contribute to the observable electron magnetic moment \cite{Bliokh}, the derivatives of the phase $\zeta_\text{fi}$ define an electric dipole moment density induced in medium by the virtual photon (see the main text).

We find
\bea
& \displaystyle \bar{u}_{\varepsilon'\lambda'}^{(\sigma')}\,{\bm \gamma}\, u_{\varepsilon\lambda}^{(\sigma)} = \left(2\lambda \sqrt{\varepsilon -m} \sqrt{\varepsilon' + m} + 2\lambda' \sqrt{\varepsilon' - m} \sqrt{\varepsilon + m}\right) \left(\omega^{(\sigma')}\right)^{\dagger} {\bm \sigma} \omega^{(\sigma)} = \cr
& \displaystyle = \left(2\lambda \sqrt{\varepsilon -m} \sqrt{\varepsilon' + m} + 2\lambda' \sqrt{\varepsilon' - m} \sqrt{\varepsilon + m}\right) 2\sigma \left({\bm \chi}^{(0)} \, \delta_{\sigma\sigma'} - {\bm \chi}^{(2\sigma)} \sqrt{2}\, \delta_{\sigma,-\sigma'}\right).
\eea
and then 
\bea
& \displaystyle \bar{u}_{\varepsilon'\lambda'}^{(\sigma')}\,\left({\bm \chi}^{(\sigma_{\gamma})}\right)^*\cdot{\bm \gamma}\, u_{\varepsilon\lambda}^{(\sigma)} = \left(2\lambda \sqrt{\varepsilon -m} \sqrt{\varepsilon' + m} + 2\lambda' \sqrt{\varepsilon' - m} \sqrt{\varepsilon + m}\right) 2\sigma \left(\delta_{\sigma_{\gamma}0} \, \delta_{\sigma\sigma'} - \sqrt{2}\,\delta_{\sigma_{\gamma},2\sigma}\,  \delta_{\sigma,-\sigma'}\right).
\eea
Summing over $\sigma,\sigma',\sigma_{\gamma}$, we notice that only the terms obeying $\sigma = \sigma'+\sigma_{\gamma}$ contribute, and so there are four of them
\bea
 & \displaystyle M_\text{fi} = g_{\lambda\lambda'}\sum\limits_{\sigma,\sigma',\sigma_{\gamma}}\delta_{\sigma, \sigma'+\sigma_{\gamma}}\, M_\text{fi}^{(\sigma\sigma'\sigma_{\gamma})}\, e^{i\zeta_\text{fi}^{(\sigma\sigma'\sigma_{\gamma})}} = g_{\lambda\lambda'}\Bigg(M_\text{fi}^{(\frac{1}{2},-\frac{1}{2},1)}\, e^{i\zeta_\text{fi}^{(\frac{1}{2},-\frac{1}{2},1)}} + \cr & \displaystyle + 
M_\text{fi}^{(\frac{1}{2},\frac{1}{2},0)}\, e^{i\zeta_\text{fi}^{(\frac{1}{2},\frac{1}{2},0)}} + M_\text{fi}^{(-\frac{1}{2},\frac{1}{2},-1)}\, e^{i\zeta_\text{fi}^{(-\frac{1}{2},\frac{1}{2},-1)}} + M_\text{fi}^{(-\frac{1}{2},-\frac{1}{2},0)}\, e^{i\zeta_\text{fi}^{(-\frac{1}{2},-\frac{1}{2},0)}}\Bigg),
\eea
where 
\bea
&& \displaystyle g_{\lambda\lambda'} = \sqrt{4\pi\alpha_{em}}\,\left(2\lambda \sqrt{\varepsilon -m} \sqrt{\varepsilon' + m} + 2\lambda' \sqrt{\varepsilon' - m} \sqrt{\varepsilon + m}\right),
\eea
and the helicity amplitudes, which are real but not necessarily positive, are
\bea 
&& \displaystyle M_\text{fi}^{(\frac{1}{2},-\frac{1}{2},1)} = \sqrt{2}\,\,  d^{(1/2)}_{1/2,\lambda}(\theta)\, d^{(1/2)}_{-1/2,\lambda'}(\theta')\, d^{(1)}_{1\lambda_{\gamma}}(\theta_{\gamma}),\quad \zeta_\text{fi}^{(\frac{1}{2},-\frac{1}{2},1)} = -\frac{1}{2}(\phi+\phi') + \phi_{\gamma},\cr 
&& \displaystyle 
M_\text{fi}^{(\frac{1}{2},\frac{1}{2},0)} = - d^{(1/2)}_{1/2,\lambda}(\theta)\, d^{(1/2)}_{1/2,\lambda'}(\theta')\, d^{(1)}_{0\lambda_{\gamma}}(\theta_{\gamma}),\quad 
\zeta_\text{fi}^{(\frac{1}{2},\frac{1}{2},0)} = \frac{1}{2}(\phi'-\phi), \cr 
&& \displaystyle
M_\text{fi}^{(-\frac{1}{2},\frac{1}{2},-1)} = -\sqrt{2}\,\, d^{(1/2)}_{-1/2,\lambda}(\theta)\, d^{(1/2)}_{1/2,\lambda'}(\theta')\, d^{(1)}_{-1\lambda_{\gamma}}(\theta_{\gamma}),\quad \zeta_\text{fi}^{(-\frac{1}{2},\frac{1}{2},-1)} =  -\zeta_\text{fi}^{(\frac{1}{2},-\frac{1}{2},1)},\cr 
&& \displaystyle
M_\text{fi}^{(-\frac{1}{2},-\frac{1}{2},0)} = d^{(1/2)}_{-1/2,\lambda}(\theta)\, d^{(1/2)}_{-1/2,\lambda'}(\theta')\, d^{(1)}_{0\lambda_{\gamma}}(\theta_{\gamma}),\quad
\zeta_\text{fi}^{(-\frac{1}{2},-\frac{1}{2},0)} = -\zeta_\text{fi}^{(\frac{1}{2},\frac{1}{2},0)}.
\eea

\begin{figure}\center
\includegraphics[scale=0.6] {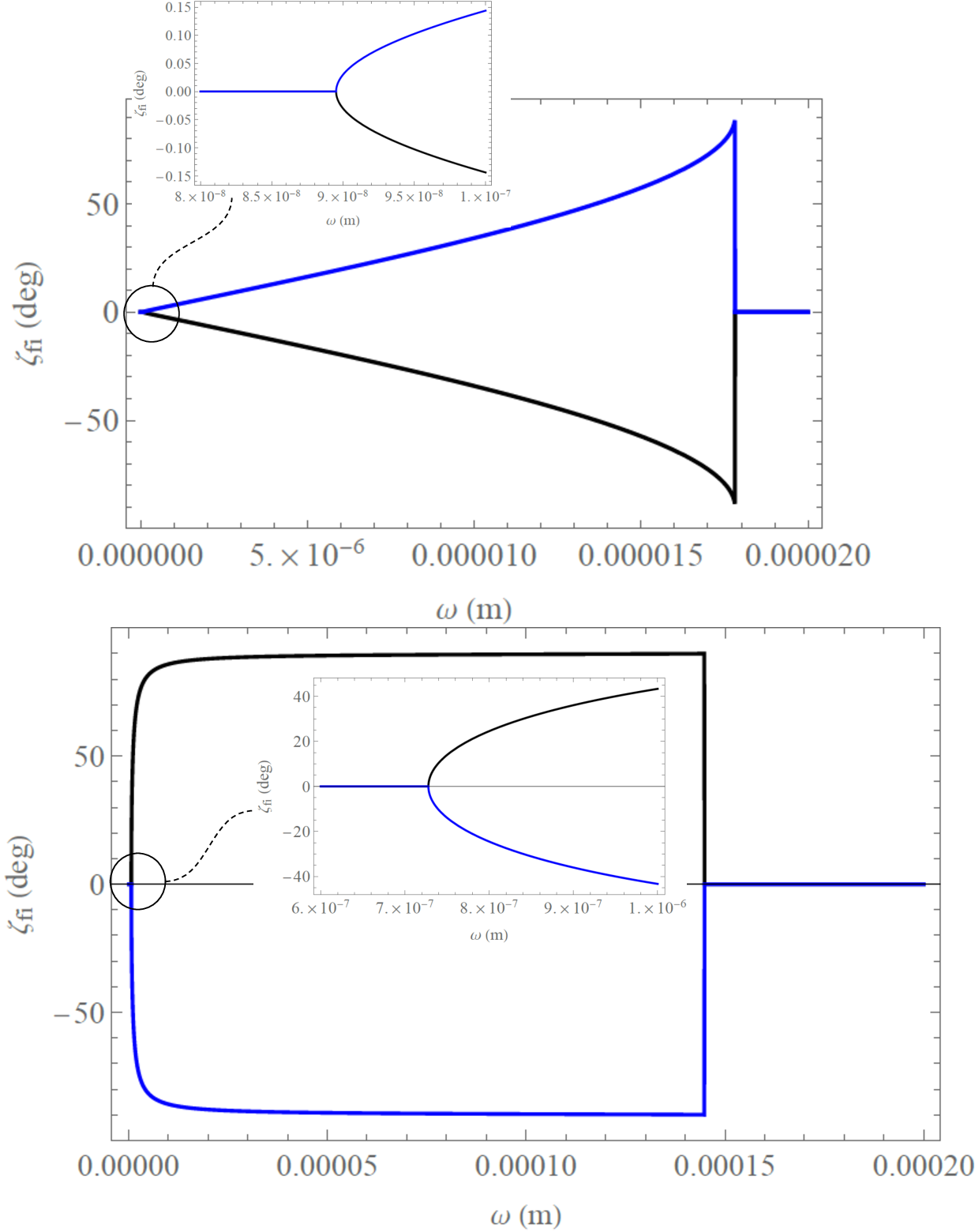}
    \caption{The phase of the scattering amplitude $\zeta_\text{fi} = \text{arg}\,M_\text{fi}$ from Eq.(\ref{MDz}) at the Cherenkov angle including the quantum recoil (\ref{Changle}) for $\beta = 0.999$. Black line is for the first triangle point from Eq.(\ref{delta2}), whereas the blue one is for the second point. Upper panel: $n = 1.5, p_{\perp} = 10^{-5} m, p'_{\perp} = 0.99\times p_{\perp}, p'_z = 0.99\times \beta, \lambda = \lambda'=1/2,\lambda_{\gamma}=\pm 1$. Lower panel: $n = 1.7, p_{\perp} = 10^{-4} m, p'_{\perp} = 0.99\times p_{\perp}, p'_z = 0.99\times \beta, \lambda = - \lambda'=1/2,\lambda_{\gamma}=\pm 1$. When $\lambda = - \lambda'=-1/2$, the black and blue lines swap.}
    \label{FigDz}
\end{figure}

Finally,
\bea
& \displaystyle |M_\text{fi}|^2/g^2_{\lambda\lambda'} = \sum\limits_{\sigma,\sigma',\sigma_{\gamma}}\delta_{\sigma, \sigma'+\sigma_{\gamma}}\, \left(M_\text{fi}^{(\sigma\sigma'\sigma_{\gamma})}\right)^2 + 2 M_\text{fi}^{(\frac{1}{2},-\frac{1}{2},1)} M_\text{fi}^{(\frac{1}{2},\frac{1}{2},0)}\,\cos\left(\zeta_\text{fi}^{(\frac{1}{2},-\frac{1}{2},1)} - \zeta_\text{fi}^{(\frac{1}{2},\frac{1}{2},0)} \right) +\cr & \displaystyle +  2 M_\text{fi}^{(\frac{1}{2},-\frac{1}{2},1)} M_\text{fi}^{(-\frac{1}{2},\frac{1}{2},-1)}\,\cos\left(\zeta_\text{fi}^{(\frac{1}{2},-\frac{1}{2},1)} - \zeta_\text{fi}^{(-\frac{1}{2},\frac{1}{2},-1)} \right) +\cr & \displaystyle +  2 M_\text{fi}^{(\frac{1}{2},-\frac{1}{2},1)} M_\text{fi}^{(-\frac{1}{2},-\frac{1}{2},0)}\,\cos\left(\zeta_\text{fi}^{(\frac{1}{2},-\frac{1}{2},1)} - \zeta_\text{fi}^{(-\frac{1}{2},-\frac{1}{2},0)} \right) +\cr & \displaystyle +  2 M_\text{fi}^{(\frac{1}{2},\frac{1}{2},0)} M_\text{fi}^{(-\frac{1}{2},\frac{1}{2},-1)}\,\cos\left(\zeta_\text{fi}^{(\frac{1}{2},\frac{1}{2},0)} - \zeta_\text{fi}^{(-\frac{1}{2},\frac{1}{2},-1)}\right) +\cr & \displaystyle +  2 M_\text{fi}^{(\frac{1}{2},\frac{1}{2},0)} M_\text{fi}^{(-\frac{1}{2},-\frac{1}{2},0)}\,\cos\left(\zeta_\text{fi}^{(\frac{1}{2},\frac{1}{2},0)} - \zeta_\text{fi}^{(-\frac{1}{2},-\frac{1}{2},0)}\right) +\cr 
& \displaystyle + 2 M_\text{fi}^{(-\frac{1}{2},\frac{1}{2},-1)} M_\text{fi}^{(-\frac{1}{2},-\frac{1}{2},0)}\,\cos\left(\zeta_\text{fi}^{(-\frac{1}{2},\frac{1}{2},-1)} - \zeta_\text{fi}^{(-\frac{1}{2},-\frac{1}{2},0)}\right),\cr
& \displaystyle \zeta_\text{fi} = \arctan\frac{\sum\limits_{\sigma,\sigma',\sigma_{\gamma}}\delta_{\sigma, \sigma'+\sigma_{\gamma}}\, M_\text{fi}^{(\sigma\sigma'\sigma_{\gamma})} \sin\left(\zeta_\text{fi}^{(\sigma\sigma'\sigma_{\gamma})}\right)}{\sum\limits_{\sigma,\sigma',\sigma_{\gamma}}\delta_{\sigma, \sigma'+\sigma_{\gamma}}\, M_\text{fi}^{(\sigma\sigma'\sigma_{\gamma})} \cos\left(\zeta_\text{fi}^{(\sigma\sigma'\sigma_{\gamma})}\right)},
\label{MDz}
\eea
where the sums include only four above terms obeying $\sigma = \sigma' + \sigma_{\gamma}$.

On the triangle point from Eq.(\ref{delta2}) $\phi = \phi'+\alpha, \phi_{\gamma} = \phi'+\alpha+\gamma$, we have 
\bea
 && \displaystyle \zeta_\text{fi}^{(\frac{1}{2},-\frac{1}{2},1)} = -\frac{1}{2}(\phi+\phi') + \phi_{\gamma} \to \gamma +\alpha/2,\cr 
&& \displaystyle 
\zeta_\text{fi}^{(\frac{1}{2},\frac{1}{2},0)} = \frac{1}{2}(\phi'-\phi) \to -\alpha/2, \cr 
&& \displaystyle
\zeta_\text{fi}^{(-\frac{1}{2},\frac{1}{2},-1)} = \pi+\frac{1}{2}(\phi+\phi') - \phi_{\gamma} \to -\gamma-\alpha/2,\cr 
&& \displaystyle
\zeta_\text{fi}^{(-\frac{1}{2},-\frac{1}{2},0)} = -\frac{1}{2}(\phi'-\phi) \to \alpha/2,
\eea
where in the second point $\phi = \phi'-\alpha, \phi_{\gamma} = \phi'-\alpha-\gamma$ the phases change the signs and so
\bea
\zeta_\text{fi}\Big|_{\phi = \phi'+\alpha, \phi_{\gamma} = \phi'+\alpha+\gamma} = - \zeta_\text{fi}\Big|_{\phi = \phi'-\alpha, \phi_{\gamma} = \phi'-\alpha-\gamma},
\label{zetapm}
\eea
whereas
\bea
|M_\text{fi}|^2\Big|_{\phi = \phi'+\alpha, \phi_{\gamma} = \phi'+\alpha+\gamma} =  |M_\text{fi}|^2\Big|_{\phi = \phi'-\alpha, \phi_{\gamma} = \phi'-\alpha-\gamma}.
\eea
Thus, the phase changes the sign for two momentum configurations from Eq.(\ref{delta2}) together with the quantum shift in the photon arrival time depending on the phase derivative, whereas $|M_\text{fi}|^2$ does not. Note that the phases -- and so the matrix element -- do not depend on the azimuthal angle of the final electron momentum $\phi'$. The transverse momenta in the amplitudes must satisfy the triangle rules Eq.(\ref{trirules}). 

In Supplementary Figure~\ref{FigDz} we show that the phase is not constant even at the tree-level and it is non-vanishing in a finite region of momentum space defined by the momentum conservation law. The region of frequencies $\Delta \omega$ for which the phase and its derivative stay non-vanishing defines the magnitude of the quantum shift in the photon arrival time, discussed in the main part of the manuscript.

\bibliographystyle{h-physrev}
\renewcommand\refname{Supplementary References}
\bibliography{refs}